\documentclass[apjl]{emulateapj}
\usepackage{apjfonts}

\newcommand{\kms}{km s$^{-1}$}
\newcommand{\magsec}{mag arcsec$^{-2}$}

\newcommand{\eg}{e.g.,\ }
\newcommand{\ie}{i.e.,\ }
\newcommand{\etal}{et~al.\ }
\newcommand{\zz}{$z=0$}
\newcommand{\zt}{$z=2$}
\newcommand{\Msun}{M$_{\odot}$}
\newcommand{\Lsun}{L$_{\odot}$}
\newcommand{\rtrunc}{$r_{trunc}$}
\newcommand{\drdr}{$d \log(\rho)/d \log(r)$}
\newcommand{\ddrddr}{$d^2 \log(\rho)/d \log(r)^2$}
\newcommand{\MsunV}{M$_\odot$ pc$^{-3}$}
\newcommand{\nbody}{$N$-body}
\newcommand{\V}{$V$}
\newcommand{\muv}{$\mu_{V}$}

\begin{document}

\title{The Quantity of Intracluster Light: Comparing Theoretical and
  Observational Measurement Techniques Using Simulated Clusters}

\author{Craig S. Rudick\altaffilmark{1,2}, J. Christopher Mihos
  \altaffilmark{1}, and Cameron K. McBride \altaffilmark{3}}

\email{craig.rudick@phys.ethz.ch}

\altaffiltext{1}{Department of Astronomy, Case Western Reserve University, 10900
Euclid Ave, Cleveland, OH 44106}
\altaffiltext{2}{Institute for Astronomy, ETH
  Z\"urich, Wolfgang-Pauli-Strasse 27, 8093 Z\"urich, Switzerland}
\altaffiltext{3}{Department of Physics \& Astronomy, Vanderbilt University, 6301
  Stevenson Center, Nashville, TN 37235}

\begin{abstract}
  Using a suite of \nbody\ simulations of galaxy clusters specifically
  tailored to studying the intracluster light (ICL) component, we
  measure the quantity of ICL using a number of different methods
  previously employed in the literature for both observational and
  simulation data sets.  By measuring the ICL of the clusters using
  multiple techniques, we are able to identify systematic differences
  in how each detection method identifies the ICL.  We find that
  techniques which define the ICL solely based on the current position
  of the cluster luminosity, such as a surface brightness or local
  density threshold, tend to find less ICL than methods utilizing time
  or velocity information, including stellar particles' density
  history or binding energy.  The range of ICL fractions (the fraction
  of the clusters' total luminosity found in the ICL component) we
  measure at \zz\ across all our clusters using any definition span
  the range from $9-36\%$, and even within a single cluster different
  methods can change the measured ICL fraction by up to a factor of
  two.  Separating the cluster's central galaxy from the surrounding
  ICL component is a challenge for all ICL techniques, and because the
  ICL is centrally concentrated within the cluster, the differences in
  the measured ICL quantity between techniques are largely a
  consequence of this central galaxy/ICL separation.  We thoroughly
  explore the free parameters involved with each measurement method,
  and find that adjusting these parameters can change the measured ICL
  fraction by up to a factor of two.  The choice of ICL definition
  does not strongly affect the ICL's ability to trace the major
  features of the cluster's dynamical evolution.  While for all
  definitions the quantity of ICL tends to increase with time, the ICL
  fraction does not grow at a uniform rate, nor even monotonically
  under some definitions.  Thus, the ICL can be used as a rough
  indicator of dynamical age, where more dynamically advanced clusters
  will on average have higher ICL fractions.
\end{abstract}

\keywords{galaxies: clusters: general --- galaxies: evolution ---
galaxies : interactions --- galaxies: kinematics and dynamics ---
methods: numerical}

\section{Introduction}
Massive galaxy clusters are known to contain a luminous component
consisting of stars which reside \emph{outside} the clusters'
galaxies, often referred to as intracluster light or ICL.  While ICL
was first detected in deep photographic imaging (Zwicky 1951; Oemler
1973; Gudehus 1989), it is only with the advent of modern
CCD imaging techniques that quantitative studies of the ICL have
become possible (\eg Uson \etal 1991; Vilchez-Gomez \etal 1994;
Feldmeier \etal 2004; Mihos \etal 2005; Zibetti \etal 2005; Krick \&
Bernstein 2007; to name only a few).  For over a decade now,
increasingly precise measurements of the ICL in galaxy clusters have
shown that the ICL is an important component of the cluster's overall
optical luminosity.

Although the details of the methods through which the ICL is generated
are still debated --- and in fact there are likely a number of
processes which contribute --- the strong tidal fields generated by
galactic interactions during the dynamical evolution of the cluster
(Gnedin 2003) are likely to be a primary contributor (\eg Willman
\etal 2004; Rudick \etal 2006, hereafter R06; Murante \etal 2007;
Conroy \etal 2007; Purcell \etal 2007; Baria \etal 2009; Yang \etal
2009; Wu \& Jiang 2009; Rudick \etal 2009, hereafter R09).  Thus, the
ICL is composed of material that has been stripped into the
intracluster environment from the cluster galaxies, and is a unique
product of galaxy evolution within high density environments, such as
galaxy clusters.  As such, it can provide a wealth of information on
the dynamical history of both the cluster itself, and its constituent
galaxies.

As the study of ICL has matured, it has become increasingly apparent
that the precise nature and identity of the ICL is poorly defined.
Numerous techniques have been developed for defining and measuring the
ICL in galaxy clusters.  The commonality between all ICL detection
methods is that in all cases the ICL has very low surface brightness
and extends well beyond the traditional outer radii of the cluster
galaxies.  However, each ICL classification method makes different
assumptions about the distribution of luminosity in the cluster
galaxies, and thus systematic differences exist between them on the
precise identity, spatial distribution, and quantity of ICL.  For
instance, several authors have measured the ICL using an isophotal
limit, whereby ICL is classified as luminosity below a given surface
brightness (\eg Feldmeier \etal 2004, Zibetti \etal 2005), while
others have attempted to model and subtract the galactic light and
defining the ICL as any excess luminosity (\eg Gonzalez \etal 2005,
Seigar \etal 2007).  Because the study of ICL requires very deep, time
consuming observations (\eg Mihos \etal 2005; Krick \etal 2006), no
single unified observational program exists containing the large
number of clusters necessary for fully understanding how the ICL
varies between different galaxy clusters.  The lack of common,
well-defined measurement techniques has hampered the ability to
directly compare multiple observational studies.

Much of our understanding of the formation and evolution of the galaxy
clusters' ICL component comes from cosmological simulations of cluster
formation which focus on the ICL's dynamics.  Once again, however,
many of these studies have employed substantially different techniques
for identifying and measuring the ICL component, making direct
comparisons difficult.  The most common method is to use a binding
energy definition, whereby ICL is luminous matter which is not bound
to any individual cluster galaxy.  However, many different studies
have used significantly different variations on this basic technique
(\eg Murante \etal 2004; Sommer-Larsen \etal 2005; Purcell \etal 2007;
Dolag \etal 2010; Puchwein \etal 2010).  Furthermore, such binding
energy definitions differ fundamentally from observationally tractable
measurements of the ICL (\eg R06).  Thus, there are not only potential
systematic differences in ICL classification between studies within
the observational and simulation sub-fields, but potentially larger
systematic discrepancies likely exist between the observational and
simulation-based approaches.

With the wide variety of ICL definitions which have been used in
various studies, none can claim to be the single definitive standard.
For any study of the ICL properties, the precise definition used must
be carefully selected to match the particular science application.
For instance, while measuring the binding energy of luminous mass to
its host galaxy may not be feasible observationally, it may be an
appropriate tool for studying the mechanics of gravitational stripping
in the cluster environment, with the caveat that ICL fractions
measured in this way will not be directly comparable to observations.
Additionally, factors such as computational efficiency, repeatability,
and parameter dependence must also be considered in the choice of ICL
definition.

A more thorough understanding of the total quantity of ICL present in
clusters thus requires studying how the most common measurement
techniques function and the systematic discrepancies between the
luminosity classified as ICL in each.  In this paper we attempt to
study and clarify the relationships between many of the most commonly
used ICL classifications.  We have developed a suite of simulated
galaxy clusters specifically designed to study the dynamics of the ICL
and galactic outskirts (first described in R06 and R09).  For each of
these clusters, we measure the ICL component using a number of common
techniques, including some which are predominantly used on
observational data sets.  Section \ref{sec:simulations} describes our
simulation methods and the basic properties of each of our clusters.
In Section \ref{sec:iclmeasurements} we give a detailed description of
each ICL measurement technique we employ.  Section
\ref{sec:iclfeatures} examines the systematic differences between the
morphological features identified as ICL by each of our methods.
Section \ref{sec:discussion} contains a discussion of our results,
particularly within the context of previous studies.  Finally, Section
\ref{sec:summary} presents a summary of our main conclusions.

\section{Simulated Clusters}\label{sec:simulations}

\subsection{Simulation Techniques}
Our method for simulating galaxy clusters was first presented in R06,
and again outlined in R09.  Here, we provide a brief overview of the
process.  Over the intervening time period, however, we have made a
number of updates to the simulation techniques.  The result of this
methodological evolution is that we have two distinct sets of
simulations, which follow the same basic process, but certain aspects
of their initialization and evolution have been altered.  All
simulations begin with a large (comoving box sizes of 50-150 Mpc),
relatively low resolution cosmological dark matter only simulation
(particle mass of $\approx 5\times10^{8}$ \Msun) run from $z=50$ to
$z=0$.  Individual clusters, ranging in mass from $8\times10^{13}$ to
$6\times10^{14}$ \Msun\ (see Table \ref{tab:cluster_properties}) were
selected from these cosmological simulations at \zz\ to re-simulate at
higher resolution, including models of the luminous galaxies within
the clusters.  To do this, we trace the mass which constitutes the
\zz\ cluster dark matter halo back to its position at $z=2$.  Of
course, at $z=2$ the cluster halo has not yet formed, and the mass is
instead contained in a number of individual galaxy and group-mass
halos.  We insert our luminous galaxy models into these halos by
excising the most bound 70\% of the halos' mass and replacing it with
a galaxy model of the same mass.  For larger, group-mass halos, we
employ a halo occupation distribution, or HOD, method (Berlind \&
Weinberg 2002) which allows us to maintain our galaxy mass function by
inserting multiple galaxies into the same halo.  We use two galaxy
models, based on those described in Hernquist (1993): a disk galaxy
with a stellar component consisting of an exponential disk plus
central bulge (with a bulge-to-disk ration of 1:5), and an elliptical
galaxy with the stars in a pure Hernquist (Hernquist 1990)
distribution; additionally, both galaxy models are embedded in an
isothermal dark matter halo.  The mass resolution of the galaxy models
is fixed, such that the luminous particles have mass $1.4 \times
10^{6}$ \Msun\ with smoothing scale 280 pc, while the galactic dark
matter particles have mass $1.6 \times 10^{7}$ \Msun\ with smoothing
scale 1.4 kpc, and the mass of a galaxy is scaled by the number of
particles which resolve it.  The final galaxies are thus composed of
particles of three distinct mass resolutions (luminous particles,
galaxy dark matter, and original cluster dark matter), with a total
dark to luminous matter ratio of 10:1.  The composite halos of the
galaxies display an NFW-like structure (Navarro \etal 1996), declining
as $r^{-3}$ at large radii.  In addition to adding the high resolution
galaxy models to the cluster itself, the surrounding cosmology is
down-sampled to lower resolution by randomly selecting a subset of the
massive particles, and increasing their mass to maintain the
cosmological mass density.  The entire cosmological volume, consisting
of a high resolution galaxy cluster embedded within a low resolution
cosmology is then re-simulated for each cluster.

This collisionless, multi-resolution simulation technique allows us to
effectively focus our computational resources on the detailed
gravitational dynamics of the cluster galaxies which give rise to the
ICL.  The trade-off, of course, is that we neglect certain aspects of
galaxy and cluster evolution which may play a role in determining the
spatial distribution of luminous material in the cluster.  By omitting
hydrodynamic evolution, we cannot measure such processes as gas
accretion, star formation, ram pressure stripping, etc. which affect
the evolution of galaxies in clusters.  However, such calculations are
computationally expensive and contain significant uncertainties.  By
focusing on gravitational mechanics alone, we are able to isolate this
important mechanism and study the gravitationally driven evolution of
the cluster at higher resolution and in greater detail.  Additionally,
we have a minimum stellar mass for inserted galaxies of $5.6 \times
10^{9}$ \Msun, or about 10\% of the mass of the Milky Way.  We do not
have the resolution to properly simulate smaller galaxies, and thus we
are unable to measure their contribution the cluster's ICL content.
By inserting our luminous galaxy models at $z=2$, we may fail to place
galaxies into halos which are below our mass limit at that time, but
which later grow to exceed the limit.  However, we have tested that
reasonable changes of the insertion redshift do not have a strong
influence on the resulting ICL properties (see R06).  Despite these
caveats, our simulations have been designed to faithfully trace the
gravitational evolution of the galaxies evolving within clusters with
sufficient resolution to trace the outermost, least bound luminosity
which is most likely to be stripped to form the intracluster light.

\subsection{Cluster Properties}
There are three main ways in which our two sets of simulations differ
from one another.  First, our original simulations were run with the
\nbody\ code GADGET (Springel \etal 2001), while the updated method
employs the more recent GADGET-2 (Springel 2005).  Secondly, we
updated the cosmological parameters to reflect recent results in this
field (\eg Hinshaw \etal 2009).  The original simulations were run in
a $\Omega_{\Lambda}=0.7$, $\Omega_{M}=0.3$, $H_{0}=70$,
$\sigma_{8}=0.9$ cosmology, while the updated simulations use
$\Omega_{\Lambda}=0.75$, $\Omega_{M}=0.25$, $H_{0}=70$,
$\sigma_{8}=0.8$.  Finally, and perhaps most substantially, the HOD
parameters in the initialization scheme were modified.  In particular,
the masses of the inserted galaxies were more heavily weighted to
low-mass galaxies, by adjusting the low-mass slope of the galaxy mass
function from $\alpha \approx -1.0$ to $\alpha \approx -1.5$.

In this work we discuss the results of six simulated galaxy clusters.
Clusters C1, C2, and C3 were first presented in R06.  Each of these
was run using the original simulation methods and all have masses of
$\approx 0.9 \times 10^{14}$ \Msun.  In order to explore how the
dynamics of ICL production may depend on cluster mass, we have run
three more massive clusters, which we refer to as B22 ($2.2 \times
10^{14}$ \Msun), B35 ($3.5 \times 10^{14}$ \Msun), and B65 ($6.5
\times 10^{14}$ \Msun), each using the fully updated simulation
methods.  Additionally, in order to check the robustness of our
simulation techniques, and to ensure that there are no systematic
effects on the cluster dynamics caused by our evolving simulation
methodology, we chose to re-initialize and re-simulate cluster C2,
starting from the same initial dark matter distribution, but using the
new initialization scheme and GADGET-2 to run the simulation.  We
refer to this simulation as C2R, and in general it behaves very
similarly to the original C2, as expected.  The basic properties of
these clusters and the techniques used to simulated them are
summarized in Table \ref{tab:cluster_properties}.  For all analyses in
this paper, we define the total cluster luminosity as the luminous
material within $1.5 R_{200}$ of the cluster center of mass unless
otherwise noted.

\begin{deluxetable}{lcclcc}
  \tablecolumns{4}
  \tablewidth{0pc}
  \tablecaption{Basic Properties of Simulated Clusters
    \label{tab:cluster_properties}} \tablehead{ \colhead{Cluster} &
    \colhead{$M_{200}$}\tablenotemark{*}&
    \colhead{$R_{200}$}\tablenotemark{*} &
    \multicolumn{3}{c}{Simulation Methods}\\
    \colhead{} & \colhead{[$10^{14}$ M$_{\odot}$]} & \colhead{[Mpc]} &
    \colhead{code} & \colhead{$\alpha$} & \colhead{$\Omega_{\Lambda}$}
  } \startdata
  C1  & 0.88 & 0.93 & GADGET & $-1.0$ & $0.70$ \\
  C2  & 0.84 & 0.92 & GADGET & $-1.0$ & $0.70$ \\
  C3  & 0.87 & 0.96 & GADGET & $-1.0$ & $0.70$ \\
  C2R & 0.82 & 0.91 & GADGET-2 & $-1.5$ & $0.70$\\
  B22 & 2.2 & 1.3 & GADGET-2 & $-1.5$ & $0.75$ \\
  B35 & 3.3 & 1.5 & GADGET-2 & $-1.5$ & $0.75$ \\
  B65 & 6.5 & 1.8 & GADGET-2 & $-1.5$ & $0.75$ \\
  \enddata
  \tablenotetext{*}{$M_{200}$ and $R_{200}$ are the mass enclosed and
    radius, respectively, of a sphere with density 200 times the
    critical density.}
\end{deluxetable}

\begin{figure*}
\plotone{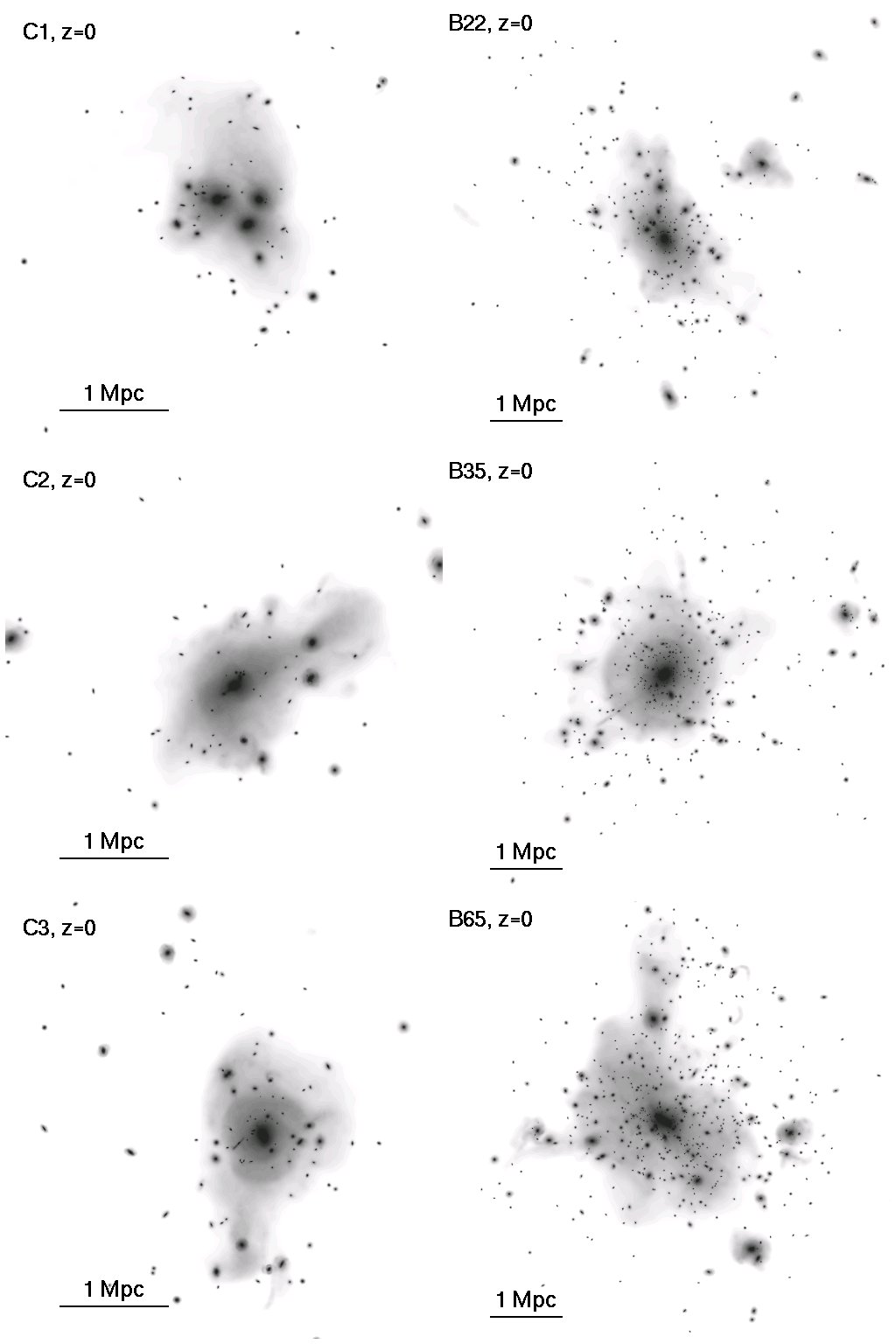}
\caption{Images of the luminous component of our six clusters at z=0.
\label{fig:clusters_z0}}
\end{figure*}

Figure \ref{fig:clusters_z0} shows an image of each of our six
clusters at \zz.  Each clusters, of course, has a unique dynamical
history, and this history is reflected in its \zz\ structure.  Cluster
C2, for instance, is very centrally concentrated with a well defined
massive central galaxy.  Because of this cluster's relaxed morphology,
and its extensive use in previous studies (\eg R06, R09), it is our
most well-studied cluster simulation.  For this reason, we often use
this cluster in the analyses below as our prime example for
demonstrating the magnitude of various effects, such as varying free
parameters in our calculations.

Another cluster of particular interest is B35, our second most massive
cluster.  While cluster B65 is nearly twice as massive, is has the
peculiar trait that at \zz\ it has recently accreted a smaller
cluster, and the two central galaxies are still in the process of
merging (the two galaxies are difficult to distinguish in Figure
\ref{fig:clusters_z0}, but are nonetheless dynamically distinct).
Therefore, in several of the analyses below we also single out B35
for more intensive scrutiny, as it is our most massive cluster which
has a well-defined central galaxy.

Structural analyses of the cluster galaxies have shown that the
simulations suffer somewhat in quality in the inner few kpc of massive
galaxies, which do not hold structural integrity over the long
simulation timescale. In the first generation of simulations (run with
GADGET), this problem manifested as material in the central regions of
galaxies artificially losing angular momentum and settling into an
overly cold central core. Because tidally stripped material comes
preferentially from the outer regions of galaxies, this defect should
not impact our ICL metrics in a significant way. In the second
generation of simulations (run with GADGET-2), this problem was fixed,
but the inner regions still suffer from continued numerical heating
that lead them to become overly diffuse. While this has the potential
to make our galaxies more susceptible to tidal stripping, we find no
evidence of these clusters being biased toward higher ICL fractions.
Again, the propensity for ICL to be stripped from the outer regions of
galaxies makes the detailed dynamics of galaxy cores a less salient
issue.

\section{ICL Measurements}\label{sec:iclmeasurements}
There is currently no single accepted measurement technique which
uniquely defines the ICL component of galaxy clusters.  In this paper,
we make quantitative measurements of the ICL in our simulated clusters
using several techniques, similar to those that have previously been
used in the literature.  Our aim is thus to better understand the
relationships between these techniques, in order to be able to compare
the results presented in studies using varied methodologies.
Moreover, any algorithm used to calculate the ICL content in clusters
will involve defining free parameters, and we therefore must also
understand how these free parameters affect the ICL measurements.  By
implementing each method on each of our clusters, we can directly
compare not only the ICL content of different clusters, but we can
study how the ICL content of an individual cluster changes when using
different techniques.

In particular, theoretical studies based on numerical simulations and
observational studies often utilize widely different ICL measurement
methods, due to the very different information accessible in each.  In
order to help bridge this gap, we have implemented three main ICL
measurement techniques.  Two of these rely on three-dimensional data
which are not observationally tractable, and thus are only measurable
in simulation data.  The third technique, however, is based on ``mock
imaging'' of our simulated clusters, and is designed to be
comparable to observational measures used on clusters in the local
universe.  In the sections below, we describe each measurement
technique, and systematically explore the free parameters on which
each depends, allowing us to examine the results of each technique on
our six simulated clusters.  The primary metric through which we
quantify the ICL content of the simulated clusters, no matter which
ICL detection method we employ, is the fraction of the cluster's
luminosity that is found in the ICL component, referred to as the ICL
fraction.

\subsection{Binding Energy}\label{sec:iclbindingenergy}
One of the most appealing and commonly used definitions of ICL is that
it is composed of stars which are gravitationally bound to the cluster
potential, but not to any individual galaxy within the cluster.  To
become unbound, these stars must have been acted upon by some external
force.  The intense tidal fields experienced by galaxies during their
evolution in a massive cluster potential provide this force, making
the ICL a unique product of galactic evolution within such dense
environments.

While using the binding energy of stars to define the ICL is appealing
from a theoretical perspective, in practice it is quite difficult to
uniquely define the potentials of individual galaxies within the
cluster.  Essentially, this is the same problem as identifying
gravitationally self-bound sub-halos within cosmological simulations.
In fact, several previous studies have used cosmological halo
detection algorithms such as SKID or SUBFIND (Murante \etal 2004;
Willman \etal 2005; Dolag \etal 2010) to define self-bound galaxies
within the cluster and calculate the binding energy of luminous
particles to these galaxies.  However, there is not a single accepted
technique which can uniquely define these sub-halos (see Maciejewski
\etal 2009 for a discussion).  Moreover, our multi-resolution
simulation approach adds an additional layer of complexity to this
already poorly defined problem.  We have therefore not implemented
such a sub-halo technique, but designed an alternative gravitational
binding energy algorithm.

In order to measure the binding energy of stellar particles relative
to the cluster galaxies, we have developed and implemented a technique
whereby for each galaxy we measure a spherical mass density
distribution with a fixed truncation radius, \rtrunc.  From such a
distribution, it is quite simple to calculate the gravitational
potential energy as a function of radius from the galaxy:
\begin{equation} \label{eqn:ebind}
\Phi(r) = -4 \pi G \left( \frac{1}{r} \int_0^r \rho(r')r'^2 \, dr' +
\int_r^{r_{trunc}} \rho(r'){r'} \, dr' \right)\end{equation} (Binney
\& Tremaine 2008) which allows us to efficiently measure the binding
energy of each stellar particle to each galaxy, and take the ICL to be
those stellar particles not bound to any galaxy.  We note that a
significant difference between our algorithm and most cosmological
sub-halo algorithms is that we do not uniquely attribute each massive
simulation particle to a particular galaxy --- \ie a single massive
particle may contribute to the binding energy of multiple galaxies ---
nor do we uniquely assign luminosity to particular galaxies, but a
luminous particle may be gravitationally bound to multiple galaxies.

In the Appendix, we provide a detailed description of our binding
energy algorithm.  Briefly, we begin by identifying each luminous
galaxy in the cluster, and measuring its spherical mass density
profile.  From this profile, we define the truncation radius to be the
point where the galaxy's density profile flattens out and is dominated
by the local background cluster density.  With this complete density
profile, we then determine for each luminous particle whether it is
bound or unbound to the galaxy.  This procedure is repeated for all of
the cluster's galaxies, and those particles which are bound to no
galaxy are identified as ICL.  While this process is straightforward
for the vast majority of cluster galaxies, a major challenge for the
algorithm is that the mass density profile of a cluster's central
galaxy is indistinguishable from the overall cluster mass density
profile.  In fact, the separation of the central galaxy from the
surrounding ICL component has long been a major issue confronting many
binding energy-based ICL definitions (\eg Murante \etal 2004;
Sommer-Larsen \etal 2005; Dolag \etal 2010).  Our particular
resolution to this issue, in which we simply define a fixed truncation
radius of 100 kpc for the central galaxies, is discussed in detail in
the Appendix.

\subsubsection{Binding Energy Results}
\begin{deluxetable*}{lcccc}
  \tablecolumns{5}
  \tablewidth{0pc}
  \tablecaption{ICL fractions of our six clusters at $z=0$, using default
  parameters for each measurement method.
  \label{tab:icl}} \tablehead{
  \colhead{Cluster} & \multicolumn{4}{c}{ICL Fraction}\\
  \colhead{} & \colhead {Binding Energy} & \colhead{Instantaneous
    Density} & \colhead{Density History} & \colhead{Surface
    Brightness} } \startdata
C1  & 10.5\% &  9.3\% & 14.8\% &  9.6\%\\
C2  & 24.5\% & 14.6\% & 20.9\% & 13.4\%\\
C3  & 19.7\% & 12.1\% & 18.5\% & 11.3\%\\
B22 & 19.0\% &  9.7\% & 15.5\% & 10.2\%\\
B35 & 26.3\%, 36\%\tablenotemark{*} & 11.1\% & 18.4\% & 10.4\%\\
B65 & 15.8\% & 10.5\% & 15.2\% & 10.8\%
  \enddata
\tablenotetext{*}{using the kinematic separation of ICL from the central galaxy}
\end{deluxetable*}
 The ICL fractions which we obtain using
this binding energy algorithm on our clusters at \zz\ are shown in Table
\ref{tab:icl}.  Because of the difficulties in defining the binding
energy of the many central group galaxies found at higher redshifts (a
situation analogous to the clusters' central galaxies at low
redshift), we have not attempted to implement our binding energy
algorithm at earlier timepoints during the clusters' evolution.
However, as a result of our galaxy initialization scheme, at \zt\
every luminous particle is bound to its host galaxy.

\subsubsection{Kinematic Separation of the Central Galaxy and
  ICL} \label{sec:kinematicseparation} While the mass density profiles
of the central galaxies do not show any features with which to
distinguish them from the surrounding ICL, Dolag \etal (2010) suggest
that the two populations may be separated on the basis of their
kinematics.  Specifically, they find that the velocity distribution of
the central galaxy plus ICL component is well fit by the superposition
of two Maxwellian distributions, each with a different characteristic
velocity dispersion.  They propose that the two Maxwellian
distributions correspond to the two luminous components --- the
central galaxy and the ICL.

\begin{figure}
  \plotone{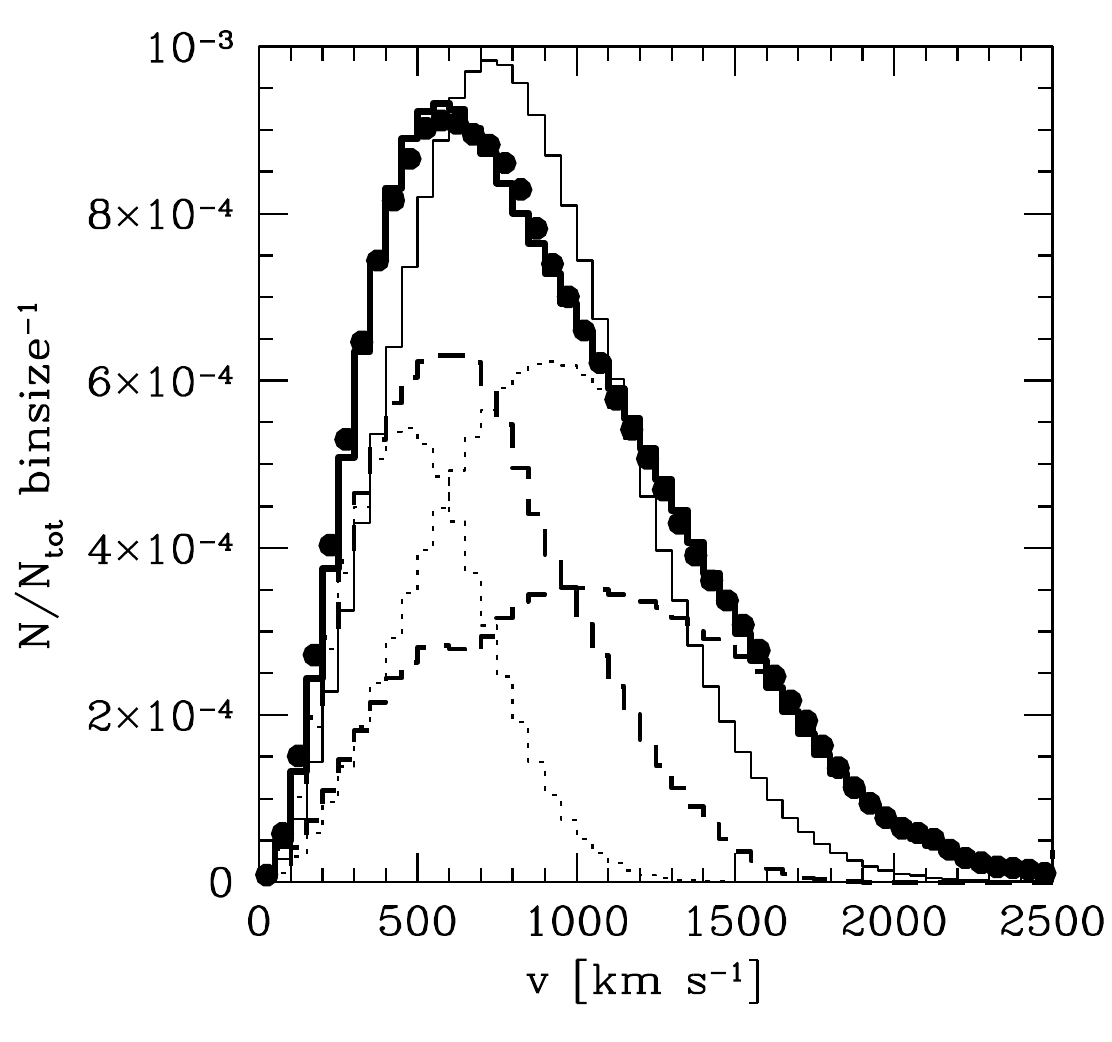}
  \caption{\emph{Solid black circles}: velocity histogram of luminous
    particles in the inner 500 kpc of cluster B35.  \emph{Thin solid
      line}: best fitting Maxwellian curve to the velocity data.
    \emph{Thick solid line}: best fitting double-Maxwellian curve to
    the velocity data.  \emph{Dotted lines}: the two components (each
    a single Maxwellian curve) of the best fitting double-Maxwellian,
    with velocity dispersions of 460 \kms\ and 920 \kms, respectively.
    \emph{Dashed lines}: velocity histograms of the bound (lower
    velocity peak) and unbound (higher velocity peak) luminous
    particles in the inner 500 kpc of the
    cluster.  \label{fig:doublemaxwellian}}
\end{figure}

As this technique has the potential to resolve the issues described
above and in the Appendix inherent in separating the central galaxy
from the ICL, we have implemented a similar analysis on our simulated
clusters.  We use cluster B35 at \zz\ to demonstrate our results, since
this cluster is our most massive cluster which has a well defined
central galaxy at \zz, and is thus likely to show the clearest
signatures of a distinction between central galaxy and ICL.  For our
lowest mass clusters, we find the Maxwellian fits to be somewhat less
robust, as the velocity dispersions of the two components are more
similar and thus more difficult to separate.

Figure \ref{fig:doublemaxwellian} shows the velocity histogram of all
particles in the inner 500 kpc of the cluster belonging to the central
galaxy plus ICL components (i.e., particles not bound to any satellite
galaxy).  We do find that a double-Maxwellian distribution is a good
fit to the velocity distribution, and the two Maxwellian components of
the fit have velocity dispersions of $460$ and $920$ \kms,
respectively, in line with those expected for a cluster of this mass
from Dolag \etal (2010).  Integrating the two fitted Maxwellian
curves, we find that the ICL component (the Maxwellian curve with the
larger velocity dispersion) contains 2.3 times as much mass as the
central galaxy.  Given that $52.2\%$ of the cluster's luminosity is
found in the central galaxy plus ICL components, this implies an ICL
fraction for the cluster of $36\%$, in line with the average ICL
fraction of $33\%$ found by Dolag \etal (2010).

In addition to simply measuring the mass of the central galaxy and ICL
using Maxwellian fits, Dolag \etal (2010) also separate the components
on a particle-by-particle basis. They do so by adjusting the
parameters of their binding energy algorithm until the velocity
distributions of each of the bound and unbound components each matches
one of the two components of the overall double-Maxwellian fit.  We
are able to achieve a similar result by reducing the truncation radius
of the central galaxy to $\approx 50$ kpc.  This value, however, is
small compared to the truncation radii of many of the largest
satellite galaxies, which reach 100-200 kpc, and implies a central
galaxy which is rather small compared to central cluster galaxies
observed in the local universe (\eg Janowiecki \etal 2010).  Thus,
while we are essentially able to match the results of Dolag \etal
(2010), we choose not to use this method as our default means of
separating the ICL from the central cluster galaxy, and continue to
use our 100 kpc truncation radius as our preferred option.  However,
it is clear that this kinematic separation technique pioneered by
Dolag \etal (2010) will result in systematically smaller central
cluster galaxies, and thus a larger and more centrally concentrated
ICL component, the implications of which will be discussed in the
following sections.

\begin{figure*} 
  \plotone{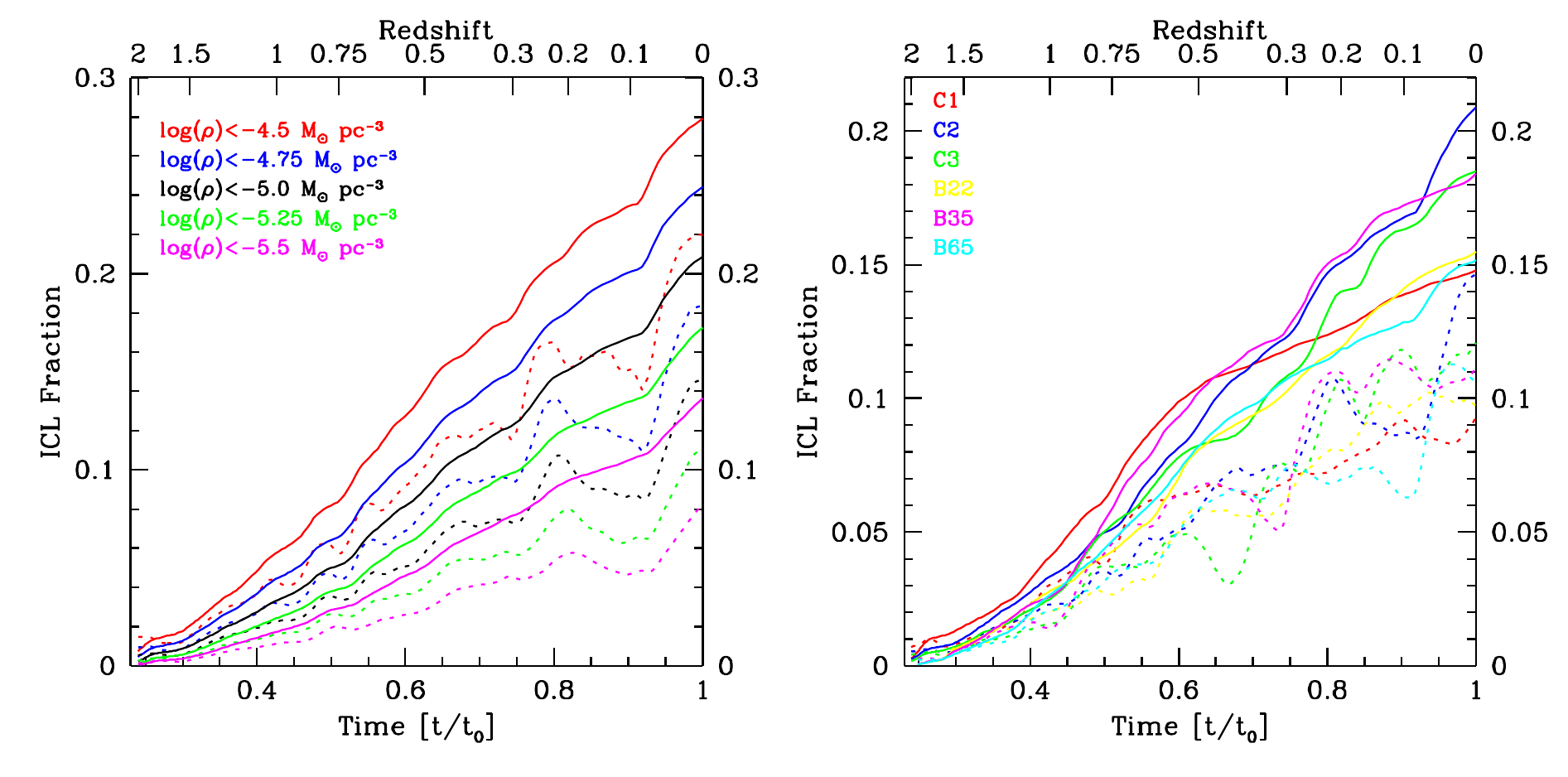}
  \caption{The ICL fraction measured by both the instantaneous density
    (dotted lines) and the density history (solid lines) methods as a
    function of evolutionary time.  The bottom axis is in units of
    $t/t_{0}$, where $t_{0}$ is the \zz\ age of the universe, while the
    top axis shows the corresponding redshift.  \emph{Left:} Cluster
    C2, using several density threshold values.  \emph{Right:} All clusters,
    using a threshold density value of $\rho \le 10^{-5.0}$ \MsunV.
    Note that since the different clusters were
    run using slightly different cosmologies, thus marginally altering the
$t/t_{0}$ : $z$ relation, the redshifts shown correctly align with the
$t/t_{0}$ axis for the $\Omega_{\Lambda}=0.7$ cosmology, and align
approximately for the $\Omega_{\Lambda}=0.75$ simulations.
\label{fig:pdenicl}}
\end{figure*}

\begin{figure*}
  \plotone{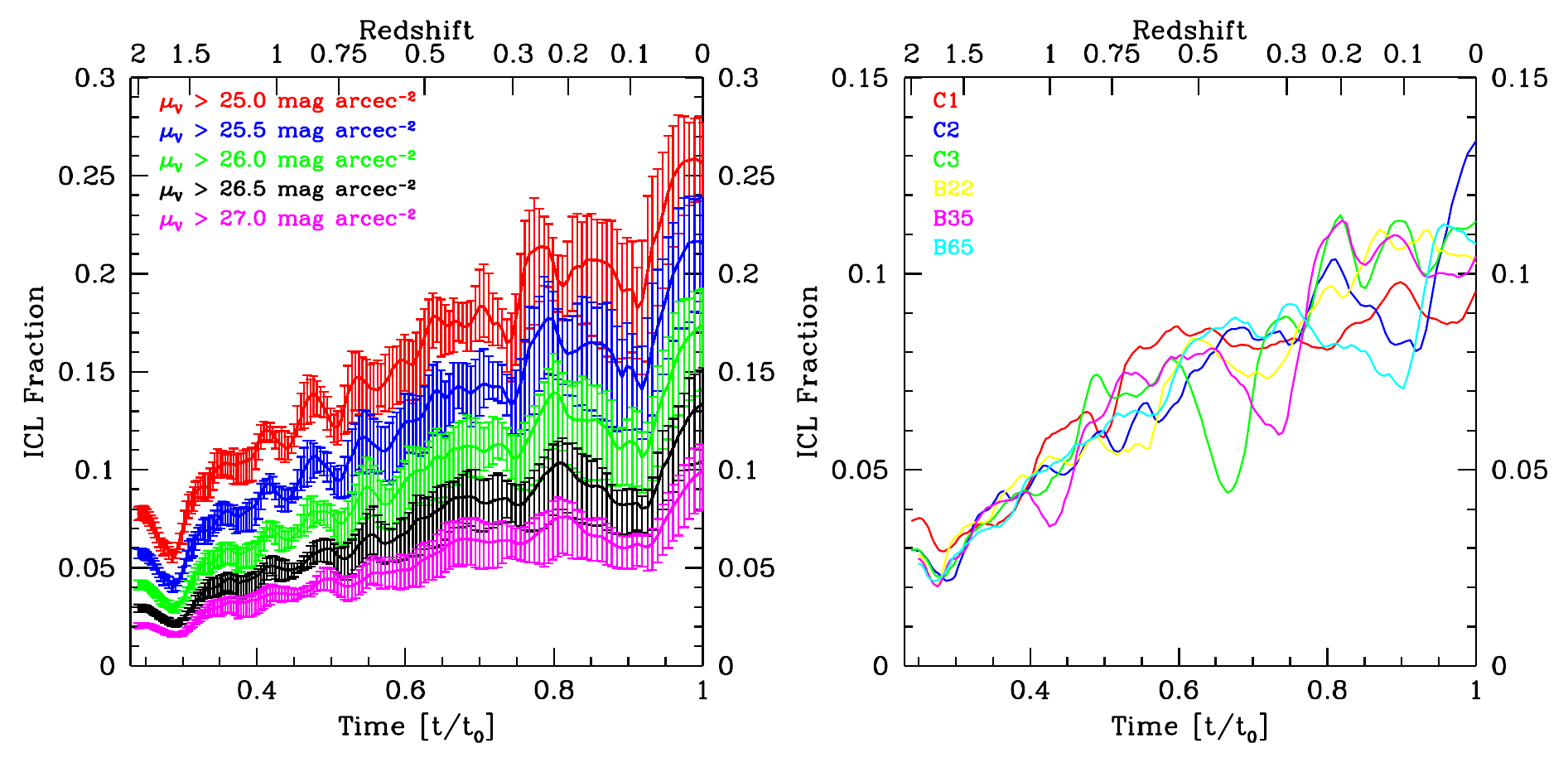}
  \caption{The ICL fraction measured by the surface brightness
    threshold definition as a function of evolutionary time. The
    bottom axis is in units of $t/t_{0}$, where $t_{0}$ is the \zz\
    age of the universe, while the top axis shows the corresponding
    redshift.  \emph{Left:} Cluster C2, using several different
    surface brightness threshold values.  The error bars show the minimum
    and maximum value found for any of the nine viewing angles, and
    the solid line follows the mean from all viewing angles.
    \emph{Right:} All clusters, using a threshold surface brightness
    value of 26.5 \magsec.  Note that since the different clusters
    were run using slightly different cosmologies, thus marginally
    altering the $t/t_{0}$ : $z$ relation, the redshifts shown
    correctly align with the $t/t_{0}$ axis for the
    $\Omega_{\Lambda}=0.7$ cosmology, and align approximately for the
    $\Omega_{\Lambda}=0.75$ simulations.
    \label{fig:sbicl}}
\end{figure*}

\subsection{Density} \label{sec:iclden}
Another method for separating the ICL from galactic luminosity in
simulation data is to define the ICL as luminous particles with low
three-dimensional density.  While this definition does not measure the
binding state of the particles to the galaxies, it is effective at
selecting only the most isolated luminosity as ICL.  We first
presented our method for implementing a density-based ICL definition
in R09.  We reiterate the method here and apply it to all of our
simulated clusters.

We calculate the density of each luminous particle as as the mass
density within a sphere of radius equal to the distance to the
particle's 100th nearest neighbor.  Testing has shown that the density
measured in this way is nearly identical to that found using the 400th
nearest neighbor, as well as other density-estimation techniques.

The simplest possible density-based ICL selection criterion is to
simply define a threshold density value below which particles are
defined to be ICL.  As in R09, our preferred value for the density
threshold is $\rho = 10^{-5.0}$ \MsunV, as this limit approximately
delineates a qualitative transition from smooth galaxy profiles to
irregularly shaped ICL features.  This effect is demonstrated in
Figure 2 from R09, which shows several galaxies in our simulations and
color-codes the luminous particles by their density.  However, the
dotted lines in the left panel of Figure \ref{fig:pdenicl} show the ICL fraction as
function of time using several different values of the density
threshold.  A key feature of this figure is that while the
curves show a great deal of structure, the different curves for the
various ICL threshold values are essentially scaled versions of one
another.  That is, the major evolutionary features of the cluster
which cause this structure (see Section \ref{sec:iclfeatures}) are
equally well represented no matter which threshold value is used, and
the different curves do not reveal unique information about the
cluster's evolutionary history.  There is, however, a small but
significant trend where the specific features of the cluster's ICL
evolution tend to occur at slightly later times for lower density ICL
threshold levels.  This result is also apparent in the surface
brightness measurements found in Section \ref{sec:iclsb}, and is
discussed in that Section.

The simple instantaneous density ICL detection method, however, is not
able to identify ICL particles which may be located in transient
high-density environments, or ICL which has migrated to near the
cluster center, where the sheer number of ICL particles may in fact
increase the local densities to above the threshold value (see Section
\ref{sec:iclfeatures}).  In order to identify such particles as ICL,
we have developed a slightly more advanced density-based criterion
which utilizes the density history of the luminous particles.  In this
method, particles which are moved to low density to become ICL
\emph{remain} classified as ICL particles, no matter their future
evolution.  However, in order to minimize the number of spurious ICL
detections due to luminous particles on highly radial orbits around
their host galaxies, we have introduced a second free parameter, a
minimum consecutive time period for which particles must be below the
threshold density value in order to be classified as ICL.  The value
of this minimum time period must balance the need to minimize spurious
detections while also not failing to detect legitimate ICL particles
which only stay at very low density for a short time, especially those
which are stripped very near to the high density center of the
cluster; our preferred time period value is 200 Myr.  For cluster C2
increasing this time period to 500 Myr causes the ICL fraction at \zz\
to decrease by 3\%, while removing the time minimum time period
restriction causes the ICL fraction to increase by 2\%.

The solid lines of Figure \ref{fig:pdenicl} show the ICL fraction
using this density history ICL definition.  A major difference between
this and the instantaneous density-based ICL techniques is that with
the density history it is, by definition, impossible for the ICL
fraction to decrease over any time interval.  Thus, for any value of
the density threshold, the density history yields a higher value for
the ICL fraction than the instantaneous density, except at the
earliest times.  For our preferred value of the density threshold,
$\rho \le 10^{-5.0}$ \MsunV, the difference in the ICL fraction at
\zz\ between the two density-based methods is 6.2\% of the cluster
luminosity.

The ICL fraction evolution for all of our simulated clusters, using
both density-based ICL detection methods, is shown in the right panel
of Figure \ref{fig:pdenicl} and the \zz\ values are recorded in Table
\ref{tab:icl}, using our preferred density threshold value of $\rho
\le 10^{-5.0}$ \MsunV.  While each cluster shows its own unique
pattern of ICL production corresponding to its particular dynamical
history, all the clusters show a similar trend of increasing ICL
fraction with time.  For most of their histories, all of the clusters'
ICL fractions lie within a fairly narrow range of a few percent.  At
\zz, the smallest ICL fractions belong to C1, at 9.2\% and 14.7\% for
the instantaneous density and density history methods, respectively,
while the largest ICL fraction belong to C2, with 14.6\% and 20.8\% of
the cluster luminosity in the ICL component under the two definitions.
Thus, we see that the production of ICL by moving luminous material
from high to low local densities is a common feature of clusters
evolution, and that the ICL fraction is correlated to the dynamical
age of the cluster, although there are variations due to the specific
dynamics of each cluster.

\subsection{Surface Brightness} \label{sec:iclsb}
In order to be able to directly compare the results of our simulated
clusters to observational studies of ICL, we have generated surface
brightness maps, or ``mock images'' of our clusters.  We first
introduced this imaging method in R06, in which we measured the ICL
content of our clusters by setting a surface brightness threshold, and
defining ICL to be luminosity at surface brightnesses fainter than the
threshold.  This ICL definition is, in fact, analogous to the
instantaneous density definition presented in Section
\ref{sec:iclden}, in that surface brightness is essentially a
measurement of the two-dimensional surface density of stellar mass,
instead of the full three-dimensional volume density.  In R06 our
primary goal was to examine the dynamical mechanisms which generated
the ICL throughout the lifetime of the cluster.  Here, we focus on
understanding the ICL measurement method itself --- what morphological
features are classified as ICL, how this compares to other ICL
definitions, and how these results vary based on the exact parameters
used in the algorithm.

Our method for generating surface brightness maps from our simulated
clusters was first described in R06.  Here we reiterate the basic
procedure for convenience.  We begin by projecting the luminous
particle distribution onto two dimensions, binned into 1 kpc pixels.
In order to create a smooth mass distribution from the discrete
luminous particles, we smooth each particle using a 2-dimensional
Gaussian kernel, with the smoothing scale, $h$, proportional to the
local 3-dimensional density of each particle, $\rho$, by $h \propto
\rho^{-1/3}$, similar to an SPH-type smoothing algorithm.  The maximum
value of the smoothing scale was set to 100 kpc for the lowest density
particles, and the smoothing kernel was truncated at $4\sigma$.

From this smooth mass distribution, we create a luminosity, and thus
surface brightness, map by applying a mass-to-light ratio to convert
from stellar mass to luminosity.  However, from our collisionless
\nbody\ simulation method, we have no information on the ages or
metallicities of the stellar features.  We have therefore chosen to
apply a uniform mass-to-light ratio to all stellar particles at all
redshifts.  We use a mass-to-light ratio of 5 \Msun\ \Lsun$^{-1}$, a
characteristic value for the \V-band light of the older stellar
populations we expect to find in galaxy clusters (Williams \etal 2007,
Rudick \etal 2010).  This unchanging mass-to-light ratio isolates the dynamical
evolution of the cluster as the sole driver of the evolving surface
brightness distribution, and thus ensures that we are not conflating
dynamical evolution with stellar evolution.  One of the ramifications
of this constant mass-to-light ratio is that the simulated images of
our clusters at high redshift are not meant to match actual
observations of high redshift clusters, but instead may be interpreted
as images of the clusters at varying dynamical ages, but containing
galaxies with similarly old stellar populations.

Given our surface brightness maps of the clusters, the simplest
possible definition of ICL is to use a surface brightness cutoff to
distinguish high-surface brightness galactic light from low-surface
brightness ICL.  In R06 we chose \muv=26.5 \magsec\ as our ICL cutoff.
This surface brightness not only corresponds to the Holmberg radius
(Holmberg 1958), a commonly used metric to determine the spatial
extent of galaxies, but also marks a qualitative transition where the
ICL begins to take on a distinct morphology from the galactic light in
both these simulations, as well as in observational works (Feldmeier
\etal 2004, Mihos \etal 2005, Rudick \etal 2010).  In this work we
more thoroughly examine the effect of this surface brightness limit on
the measured quantity of ICL.  The left panel of Figure
\ref{fig:sbicl} shows the ICL fraction as a function of evolutionary
time for several different choices of the ICL surface brightness
cutoff.  One feature of this surface brightness definition, described
in detail in R06, is that for none of the surface brightness
thresholds does the ICL fraction approach zero, even at our simulation
initialization.  This effect is due to the fact that all galaxies,
even those that remain completely tidally undisturbed, have some small
amount of material in their outermost, low-surface brightness regions,
which will be classified as ICL under this definition.

Just as was seen for the density-based ICL definitions in the left
panel of Figure \ref{fig:pdenicl}, the curves for each of the ICL
surface brightness thresholds are essentially scaled versions of one
another and do not reveal unique information about the cluster's
evolutionary history.  However, also similarly to Figure
\ref{fig:pdenicl}, there is a small but significant trend where the
specific features of the cluster's ICL evolution tend to occur at
slightly later times for fainter ICL cutoff levels.  Given the
fundamental similarities between these two ICL measurement techniques,
it is unsurprising that this phenomenon manifests itself similarly in
each.  These results are in excellent agreement with the findings of
R06, which showed a similar trend when examining the ICL evolution at
\muv$\ge26.5$ and \muv$\ge30.0$ \magsec.  This time lag indicates that
while the evolutionary dynamics which drive the ICL evolution (see
Section \ref{sec:iclfeatures}) can be seen equally well using any
surface brightness cutoff level or density threshold value, features
at lower densities and surface brightnesses probe events which
occurred in the more distant past.  Thus, while at a given time the
relative ICL fractions measured using different surface brightness
cutoffs or density thresholds give some indication of the timing of
ICL production events, all are fundamentally probing the same events.

An important difference between this surface-brightness definition of
ICL versus either the binding energy or density-based ICL definitions
is that the surface brightness method does not classify individual
luminous particles in the simulations as ICL or not.  Instead, the
luminosity in each pixel is defined as ICL or galactic based on the
pixel's surface brightness.  Because the luminosity in each pixel is
the result of the smoothed distribution of luminous particles, each
pixel contains luminosity originating from many different particles,
and a single luminous particle may contribute to the luminosity in
pixels of both galactic and ICL surface brightness.  This distinction
does not prevent us from comparing the ICL fractions measured between
this and other definitions, but it does prevent us from comparing the
results of the surface brightness definition with those of the other
definitions on a particle-by-particle basis (see Section
\ref{sec:iclfeatures}).  

Because of the nature of observational metrics which project the
three-dimensional structure of the clusters onto two dimensions, the
precise viewing angle at which the clusters are observed can affect
the resulting ICL measurements.  The effects of changing the viewing
angle on the ICL fraction of cluster C2 is also shown in the left
panel of Figure \ref{fig:sbicl}.  At each timepoint, we image the
cluster from nine different viewing angles, chosen so that they are
approximately evenly spaced on a sphere, and measure the ICL fraction
of each.  The error bars show the minimum and maximum values for any
of these viewing angles, and the solid line follows the mean from all
viewing angles.  While the major evolutionary trends are similar for
all viewing angles, the absolute value of the ICL fraction varies by
$\pm \approx 2\%$ at \zz, across the different viewing angles.

Figure \ref{fig:sbicl} shows the \muv$\ge 26.5$ \magsec\ ICL
fractions, as a function of time, for all six simulated clusters, and
Table \ref{tab:icl} lists the values at \zz. Once again, these results
are very similar to those seen in Figure \ref{fig:pdenicl} for the
density-based ICL definitions.  Each cluster's ICL fraction grows as a
function of time following a unique pattern related to its dynamical
history.  However, all the clusters stay within a relatively small
range of a few percent for most of their evolution.  At \zz, the
smallest ICL fraction, again belonging to cluster C1, is 9.5\%\ while
the largest is again cluster C2 with 13.2\%.  Just as galaxy evolution
within the cluster environment inexorably moves luminous material from
high to low density, it also moves luminosity from high to low surface
brightness.

\section{Morphological Features of the ICL} \label{sec:iclfeatures}

Each of the ICL definitions presented in Section
\ref{sec:iclmeasurements} is designed to identify stellar material in
our simulated clusters that has been stripped from the individual
galaxies during their evolution within the cluster.  However, because
each does so in a unique manner, we expect that both the qualitative
features and measured quantity of ICL to vary between the different
techniques.  We stress, however, that there does not exist an
unambiguous definition of what is or should be classified as ICL and
there is thus no single criterion with which to judge the quality of
any given definition.  In this section, we aim to better understand
the morphological features which are defined as ICL under each of our
implemented definitions, and how these ICL features relate to the
clusters' evolution and dynamics by analyzing the spatial
distribution, total quantity, and time evolution of the ICL in our
simulated clusters.

\begin{figure*}
  \plotone{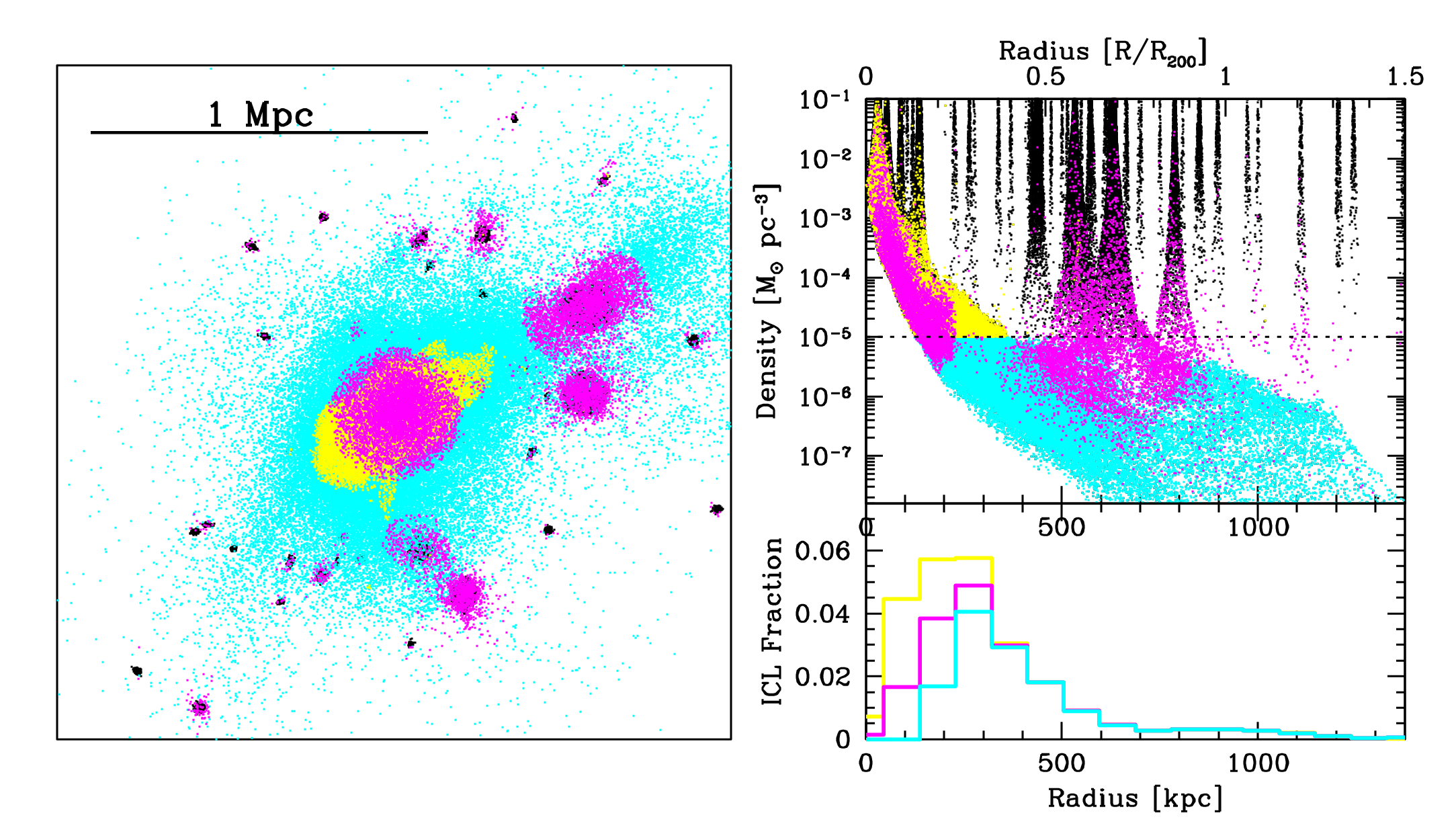}
  \caption{\emph{\bf{Left:}} The positions of luminous particles in cluster C2 at \zz.
    To increase clarity, only 3\% of the particles have been plotted.
    Even so, in certain areas, especially the cluster center,
    overcrowding is a problem and particles may be over-plotted on top
    of one another, making certain features (such as the locations of
    galaxies in the cluster core) difficult to discern.  \emph{Black}:
    not identified as ICL using any method; \emph{Cyan}: identified as
    ICL using each of the binding energy, instantaneous density, and
    density history methods; \emph{Yellow}: identified as ICL from
    binding energy only, and not from either density-based technique;
    \emph{Magenta}: identified as ICL from either or both
    density-based methods, but not from binding energy.  Particles
    identifed as ICL by the binding energy method but only one of the
    two density-based mehtods are very few in number, and are not shown
    for the sake of clarity.  These classifications do not utilize our surface brightness
    definition since it does not classify ICL on a
    particle-by-particle basis; due to the similarities between
    instantaneous density and surface brightness, however, there is an
    excellent correlation between the two definitions and, in general,
    areas of low instantaneous density will have low surface
    brightness. \emph{\bf{Right:}} The radial distribution of ICL in cluster C2 at \zz.  The
  bottom axis shows the cluster-centric radius in kpc, while the top
  axis shows the corresponding radius in terms of the cluster's
  $R_{200}$.  \emph{Top}: The cluster-centric radius of each particle
  plotted against its density.  The ICL status of the particles is
  delineated using the same colors as in the left panel.  The
  dotted line shows $\rho=10^{-5}$ \MsunV, the value of the density
  threshold. Each thin pyramidal structure at high density corresponds
  to an individual galaxy.  Note that the cluster center (R=0) is
  defined as the center of mass of the cluster, and the cluster's
  massive central galaxy sits $\approx 30$ kpc from this center.
  \emph{Bottom}: Histograms of the ICL fraction (measured as the ratio
  of ICL in the bin to total cluster luminosity) in radial bins from
  the cluster center for three of our ICL definitions: binding energy
  (yellow, filled), instantaneous density (cyan, hatched), and density history
  (magenta, hatched).
  \label{fig:iclimage}}
\end{figure*}

Much about the nature of ICL, the dynamics which produce it, and the
specific characteristics of each ICL definition, can be deduced simply
from its spatial distribution.  Figure \ref{fig:iclimage} and shows
several methods which we have used to study the spatial distribution
of ICL in cluster C2 at \zz.  The left panel of Figure
\ref{fig:iclimage} shows a map of the cluster's luminous particles,
color coded based on whether or not they are identified as ICL using
several of our techniques.  On the top right of Figure
\ref{fig:iclimage} each particle's density is plotted against its
distance from the cluster's center of mass, again color coded by ICL
status.  Finally, the bottom right of Figure \ref{fig:iclimage}
shows the radial distribution of the ICL for each of our definitions
(excluding surface brightness, since it does not identify specific
particles as ICL, as discussed in Section \ref{sec:iclsb}).

\begin{figure}
  \plotone{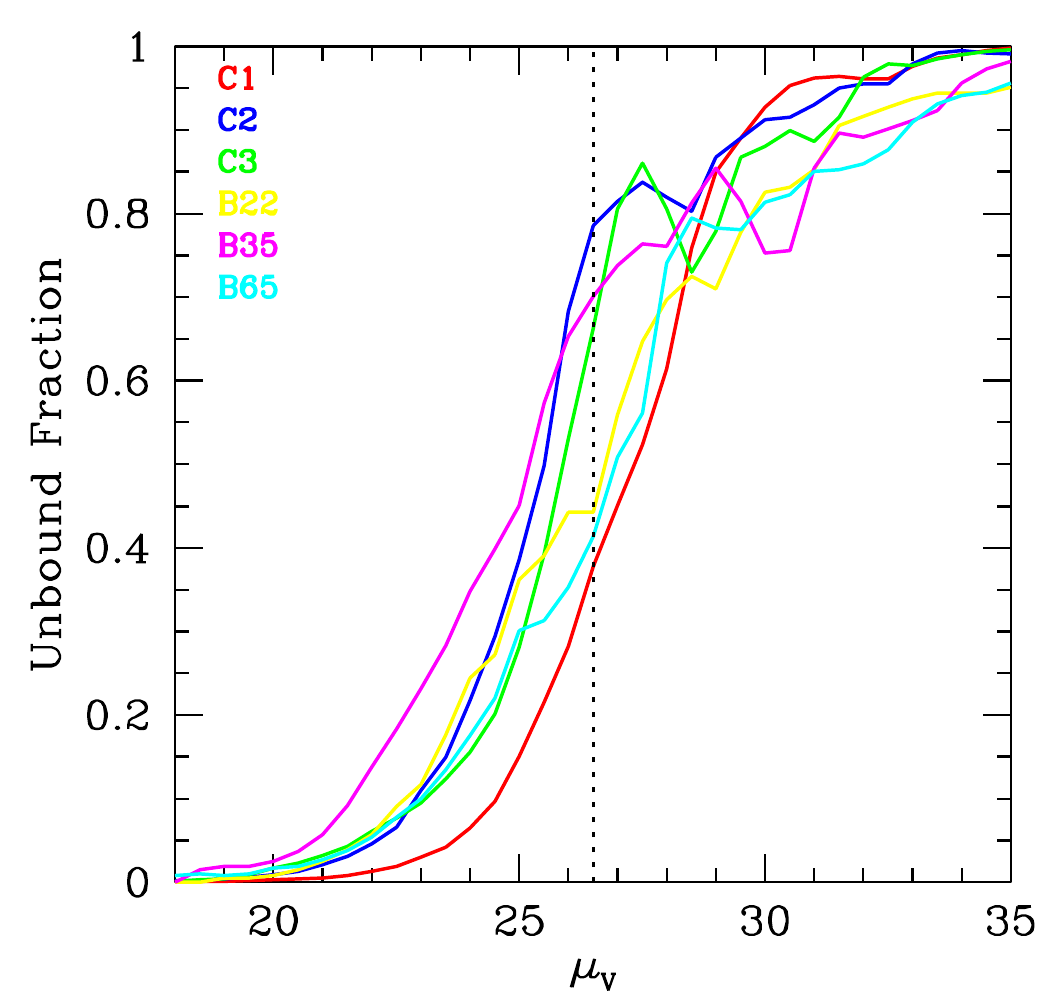}
  \caption{The fraction of luminosity at a given surface brightness
    which is contributed by unbound stars for each of our clusters.
    \label{fig:surfb_ebind}}
\end{figure}

\begin{figure}
  \plotone{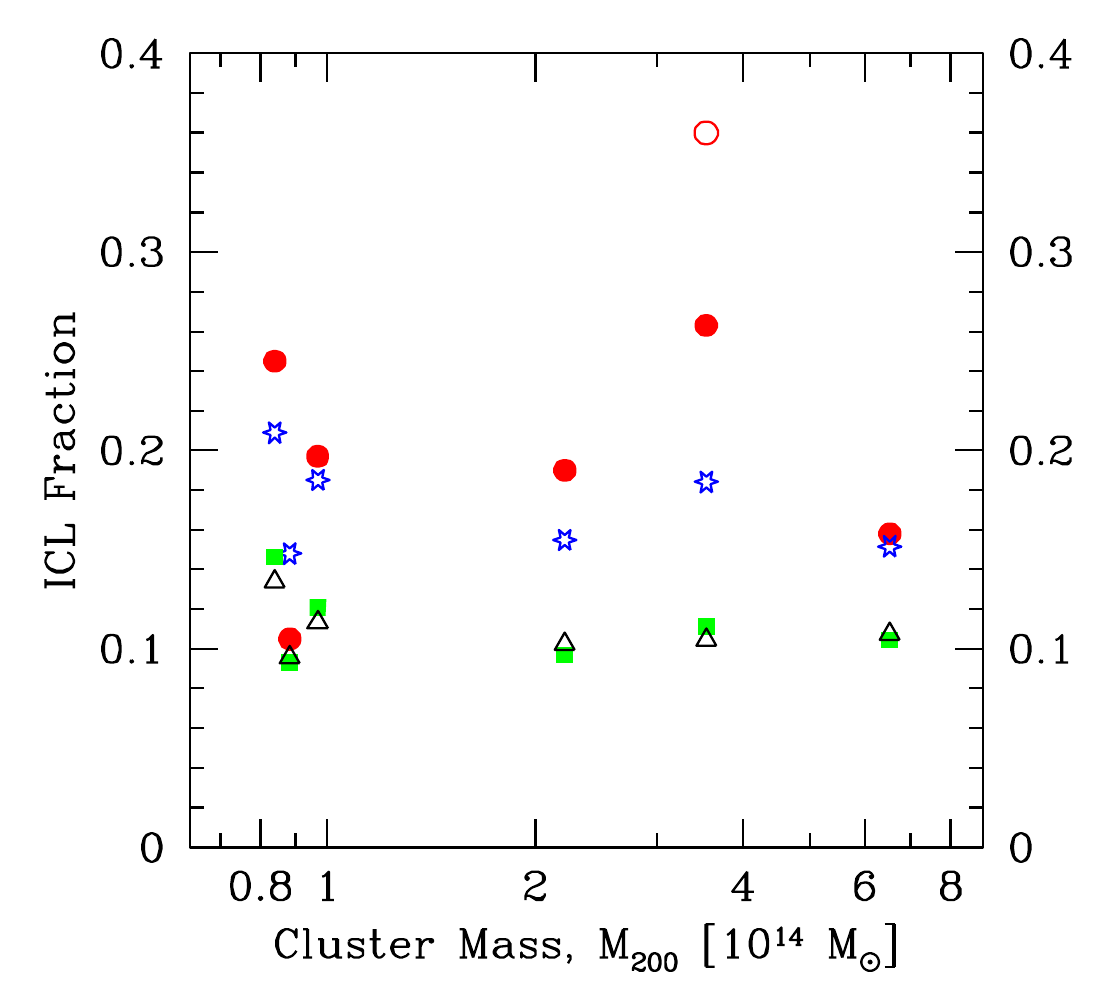}
  \caption{The ICL fraction of each cluster, using each of our four
  definitions, versus the cluster mass.  Each point type corresponds to
  a different ICL definition: binding energy (red solid circles),
  binding energy using a kinematic separation of central galaxy and
  ICL (red open circle),
  instantaneous density (black open triangles), density history (blue stars), and
  surface brightness (green solid squares).
  \label{fig:iclvmass}}
\end{figure}

From Figures \ref{fig:iclimage}, a number of key features of the ICL's
spatial distribution are apparent.  First, the cluster can be roughly
divided into two regions: an inner core where the densities of all
particles are above the density threshold used in the density-based
ICL definitions, and the outer regions where the densities can drop
below this limit.  The core is approximately ellipsoidal in shape and
centered on the cluster's central galaxy.  The main feature which
distinguishes the core is, of course, that none of the particles in
this region can be classified as ICL based on their instantaneous
density.  However, a number of particles classified as ICL based on
their density history are found in this region, and are the primary
reason that the density history produces a higher ICL fraction than
the instantaneous density definition.  These particles must have
either been stripped outside the core and migrated inward, or been
stripped early in the cluster's evolution before the core formed.
Newly stripped particles within the core are likely to remain in this
high density region --- especially when stripped from galaxies on
circular orbits at small radii --- and therefore will not be
classified as ICL based on their density history.  This is not the
case, however, for ICL identified by binding energy, since this
definition is not in any way dependent on the local density, and even
at very small radius, the intense gravitational forces are capable of
unbinding some stellar particles.  This population of small-radius,
and thus high-density, unbound particles is the major reason the
binding energy definition produces the highest ICL fraction for this
cluster.  Moreover, as noted in Section \ref{sec:kinematicseparation},
a kinematic separation of the central galaxy and ICL would only
further increase the ICL fraction by identifying many more ICL
particles at the smallest radii.

Outside of the core, the vast majority of particles tagged as
ICL in one definition are also identified in the others.  There are,
however, two notable exceptions to this trend, both particularly
apparent at the outskirts of the cluster's largest satellite galaxies.
First, a significant population of particles, especially in the inner
regions of recently created tidal streams, are classified as ICL by
instantaneous density and density history, although are bound to their
host galaxies.  This suggests that material stripped from satellite
galaxies may not immediately become unbound from its host, although
it is likely to become unbound during its subsequent evolution.
Secondly, there is a population of particles identified as ICL by
density history, although currently above the instantaneous density
threshold.  While some of these are simply particles on highly radial
orbits, others have truly been tidally stripped and then re-accreted
by the galaxy.

The majority of the discrepancies between our ICL definitions
occur in the inner core of the cluster.  Although the direct
algorithmic cause of this phenomenon is the region's high densities,
more fundamentally it points toward the fact that there is no clear
distinction between the cluster's central galaxy and its extended ICL
envelope, just as in Section \ref{sec:iclbindingenergy} we saw that
the mass density profiles of central galaxies were indistinguishable
from the overall cluster profiles.  While we identified small
differences in how the definitions identify ICL at the outskirts of
satellite galaxies, these represent only a small fraction of the
cluster's ICL for two reasons: 1) the distinction between the galaxies
and ICL is more clear for satellite galaxies and 2) the bulk of the
cluster's luminosity, and especially its potential ICL component, is
centrally concentrated.  The density versus radius plot in the top of
Figure \ref{fig:iclimage} shows a very clear locus of luminous
particles extending outward from the central galaxy.  This locus
contains the bulk of the ICL luminosity under any definition, however
there are no obvious features along the locus which can be used to
separate the light of the central galaxy from the ICL.  Under any of
our ICL definitions, the primary effect of changing the technique's
free parameters is to change the subset of particles along this locus
that is classified as ICL.  In Section \ref{sec:iclbindingenergy}, we
explicitly showed that the binding energy of luminous particles to the
central galaxy is poorly determined, and that the ICL fraction using
this definition is very heavily influenced by the precise parameters
used to define the central galaxy.  Figure \ref{fig:iclimage} also
makes clear that adjusting the free parameters of the density-based
definitions (and, by analogy, the surface brightness definition; or
see Figure 5 of R06) also primarily changes the ICL fraction by
affecting the ICL surrounding the central galaxy.

One consequence of the highly centrally-concentrated nature of the ICL
is that our measured ICL fractions tend to increase slightly when the
outer regions of the cluster are excluded.  For instance, the ICL
fraction of cluster C2 measured by binding energy within R$_{500}$ is
$28.1\%$, a modest increase from $24.5\%$ measured within our fiducial
1.5 R$_{200}$ limit.  Of all of our ICL definitions, the binding energy
method, being the most centrally concentrated, is the most affected by
this change in limiting cluster radius.  The ICL fraction measured by
density history, for example, increases from $20.9\%$ within 1.5
R$_{200}$, to only $21.9\%$ within R$_{500}$, and the other
definitions show similarly small changes.  Therefore, while the choice
of limiting cluster radius may have a small systematic influence on
the measured ICL quantities, the main results described in this work
are not highly dependent on this choice.

While we are unable to compare our density or binding energy-based ICL
definitions to our surface brightness threshold on a
particle-by-particle basis, we still wish to explore the observable
properties of the ICL identified with these methods.  In particular,
Figure \ref{fig:surfb_ebind} shows the surface brightness of the
unbound stars (i.e. ICL by binding energy) by plotting the fraction of
the cluster luminosity which is unbound at a given surface brightness
for each of our six clusters.  These measurements were made by
creating ``mock images'' of only the unbound stellar populations of
the clusters using exactly the same methodology as in Section
\ref{sec:iclsb} and comparing them to the original images.  While
there are subtle differences in the slopes of the curve for each
cluster, they all display a distinctive S-shaped trend.  At surface
brightnesses brighter than \muv$\approx 21$, an extremely small
fraction of light is contributed by unbound stars.  At fainter surface
brightnesses, however, the contribution of the unbound stars quickly
increases.  At surface brightnesses fainter than \muv\ of $26.5$, our
default surface brightness threshold for ICL, unbound stars contribute
$\gtrsim 50\%$ of the luminosity.  At the very faintest surface
brightnesses, of course, the unbound fractions converge toward unity,
as the vast majority of the stellar material at these surface
brightnesses is unbound.  These results highlight the fact that while
there is certainly a correlation where unbound material tends to be
observed at faint surface brightness, there does exist some unbound
material at relatively bright surface brightness, and vice-verse.
Inspection of the images generated using only unbound stars confirm
that the vast majority of the high-surface brightness unbound material
is centrally concentrated in the cluster core, where the projected
luminosity densities are highest.  Conversely, the bound material at
very low surface brightness comes primarily from tidal features around
galaxies outside the core of the cluster, where material can remain
loosely bound to its parent galaxy for some time before being fully
stripped by the cluster potential.  These results are in direct
analogy to the instantaneous density definition which does not
identify ICL in the innermost regions of the cluster due to the high
volume-densities of luminous particles, but does immediately classify
the material in tidal streams as ICL, thus re-confirming the
similarities between the surface brightness and instantaneous density
methods.

Given that each definition we have presented selects a different
population of luminous particles as the ICL, it is unsurprising that
the total measured quantity of ICL in the cluster is dependent on
which definition is chosen.  For instance, the fact that an
instantaneous density definition does not identify as ICL any
luminosity near the cluster core causes the ICL fraction measured
using this method to be lower than those measured using either the
density history or binding energy techniques.  Because of the innate
similarities between the instantaneous density and surface brightness,
the surface brightness-based definition of ICL displays a similar
effect.  One interpretation of this phenomenon is that the specific
morphology of a cluster may be able to ``hide'' stripped stellar
material that is spatially coincident with the cluster's galaxies from
ICL definitions which rely on position only, such as instantaneous
density and surface brightness.  In fact, this geometric effect is
also readily apparent in the surface brightness ICL definition, as it
is the primary cause of the small variance in the ICL fraction based
on the viewing angle.

To further illustrate the variance in ICL fraction that is found
between different ICL definitions, Figure \ref{fig:iclvmass} plots the
measured ICL fractions for each of our clusters at \zz, using all four
of our ICL definitions.  These results show that the basic results
discussed above in relation to cluster C2's ICL features also hold for
our other clusters.  For instance, in each case the surface brightness
and instantaneous density definitions not only the give the lowest ICL
fraction values of the four definitions, but these ICL fractions
differ by a maximum of 1.2\%, thus re-confirming the strong
similarities between these definitions.  The ICL fractions from
density history and binding energy, however, are significantly higher,
and in all but one case are higher than the ICL fraction from
instantaneous density by at least 5\%.  The one exception is the
binding energy ICL fraction calculated for cluster C1; this cluster
displays a unique morphology with three nearly equal-mass large
galaxies near the cluster center.  Such a scenario exacerbates the
difficulties in dealing with central galaxies that are inherent in
binding energy definitions, and the fact that an unusually large
fraction of the cluster's luminosity is bound to these galaxies only
further demonstrates that massive central galaxies play a unique role
in determining a cluster's ICL content.  Additionally, from the
results of Section \ref{sec:kinematicseparation} we can infer that
using a kinematic separation of the ICL from the central galaxy would
be likely to generate the highest ICL fractions for nearly every
cluster (the results of such a separation on cluster C1, however,
without a single central galaxy, remain poorly constrained).

Figure \ref{fig:iclvmass} not only plots the measured ICL fractions
for each of our clusters, but does so as a function of the cluster
mass.  We do not find any trend of increasing or decreasing ICL
fraction with cluster mass using any of our ICL definitions.  While
some studies have reported a trend where higher mass clusters tend to
have greater ICL fractions (Murante \etal 2007; Purcell \etal 2007),
other works have seen little evidence for such a trend (Dolag
\etal 2010; Puchwein \etal 2010).  Given our very limited sample size
and the wide scatter found in those studies which find a trend with
mass, we do not feel that our results are able to support or
refute either scenario.

\begin{figure}
  \plotone{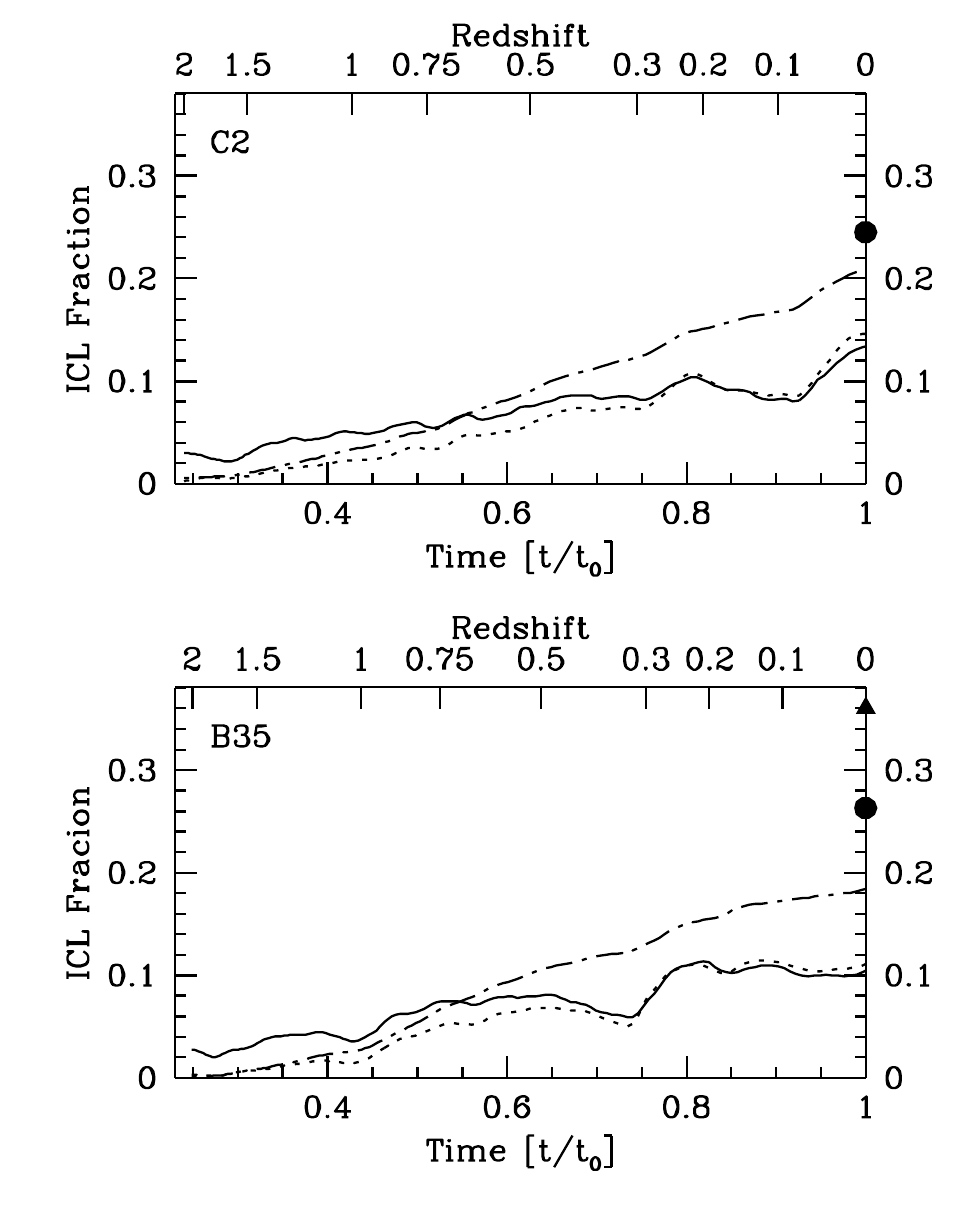}
  \caption{The ICL fraction as a function of evolutionary time, using
    each of our four ICL definitions, for clusters C2 (top) and B35
    (bottom).  Each line type corresponds to a different ICL
    definition: instantaneous density (dotted line), density history
    (dash-dot line), and surface brightness (solid line).  The binding
    energy ICL fraction is only calculated at \zz, and is shown with a
    closed circle and the binding energy variant using a kinematic
    separation of the ICL is shown at \zz\ for cluster B35 as a solid
    triangle; the ICL fraction using the binding energy definition at
    \zt\ is precisely zero for both clusters.  The bottom axis is in
    units of $t/t_{0}$, where $t_{0}$ is the \zz\ age of the universe,
    while the top axis shows the corresponding redshift.
  \label{fig:allmeasures}}
\end{figure}

While the properties of the ICL in our clusters at \zz\ provide
insight into the morphological features identified as ICL using our
various definitions, an even greater understanding of the dynamical
evolution of the clusters can be gained from the evolution of the ICL
fraction as a function of time.  Figure \ref{fig:allmeasures} shows the
ICL fraction as a function of time for both cluster C2 and B35, using
each of our definitions; the binding energy ICL fraction is shown only
at $z=0$ since it is not calculated at every timepoint for the reasons 
discussed in Section \ref{sec:iclbindingenergy}, and is exactly zero
at the cluster's initialization at $z=2$.  As expected, no matter the
definition, each of the clusters shows an increasing ICL fraction as a
function of time as a result of tidal stripping within the galaxy
cluster.  Importantly, however, the ICL fraction does not increase at
a constant rate under any ICL definition for which we have time series
data, and under the surface brightness and instantaneous density
definitions the increase is not entirely monotonic.  This is a result
that was previously discussed in both R06 and R09, where major ICL
production events were found to be correlated with specific galactic
interactions occurring within the cluster.  In Figure
\ref{fig:allmeasures} we see that these ICL production events occur at
essentially the same times, and are of similar magnitudes, under each
the different ICL definitions.  This implies that each definition is
primarily affected by the same events, and that any of these
definitions are effective at tracing the dynamical history of the
cluster.

The same variability in the cluster's evolving ICL fraction that
provides information on its dynamical history, also implies that at
any single timepoint the ICL fraction alone is not a robust indicator
of the cluster's dynamical state.  For instance, over the last
gigayear of cluster C2's evolution, the ICL fraction based on the
surface brightness definition increases by 4.3\% of the cluster's
luminosity, which is over 30\% of the total \zz\ ICL fraction of
13.4\%.  Thus, while we can expect that, on average, a cluster's ICL
fraction will increase as it dynamically ages, there is not a direct
1:1 correlation between the two.  When comparing clusters, we can
expect that, on average, clusters with higher ICL fractions will be
more dynamically advanced than clusters with lower ICL fractions.
Although the short-term variations of the ICL fraction make such
comparisons unreliable for individual clusters, it may be a metric
well-suited to studying large populations of clusters statistically,
such as that of Zibetti \etal (2005) using clusters from SDSS.
Similarly, several theoretical studies have measured the mean and
variance of the ICL fraction as a function of cluster mass using large
samples of simulated clusters (Murante \etal 2004; Purcell \etal
2007, Dolag \etal 2010).

Even if the precise ICL fraction of a cluster may have a somewhat
ambiguous relationship to its dynamics, it is clear that the evolution
of galaxies within the cluster environment invariably generates a
significant ICL component.  Using any of our ICL definition, the \zz\
ICL fractions for our clusters range from $9-36\%$.  Obviously, the
ranges within a single definition, for a single set of parameters,
are much smaller, but this wide range illustrates that point that no
matter how the ICL is defined a significant population is present.
Although this range is not meant to put definitive limits on the range
of possible ICL values in observed clusters using any ICL definition,
it demonstrates that all massive clusters at low redshift are expected
to have at least some observable ICL component, but that the majority
of the clusters' stars should still be contained within the galaxies
themselves.

\section{Discussion} \label{sec:discussion} Comparing the ICL
fractions measured for our simulated clusters to those found in the
literature using similar techniques, we generally find excellent
agreement.  For instance, Feldmeier \etal (2004) use a pure isophotal
limit to measure the ICL in a number of observed clusters, and find
ICL fractions of $7-15\%$ when using \muv$\ge26.5$, in excellent
agreement with the $9-13\%$ range we measure in our simulations.
Simulation-based measurements which use a binding energy definition of
ICL, however, tend to find somewhat higher ICL fractions.  Several
examples of such studies, each using unique variation of the binding
energy method, include: Murante \etal (2007) who measure ICL fractions
of $\approx 10-30\%$; Sommer-Larsen \etal (2005) who find $\approx
20-40\%$ of the cluster luminosity in the ICL; and Purcell \etal
(2007) who report ICL fractions ranging from $\approx 20-30\%$ for
massive clusters.  Again, these match closely to the $10-26\%$ ICL
fraction range that we find using our binding energy technique, and
fit with our expectation that a binding energy definition produces a
higher ICL fraction than an isophotal limit.  Additionally, the
variant of the binding energy technique which uses kinematics to
separate the central galaxy from the surrounding ICL, produces
potentially the highest ICL fractions, with a range of $\approx
20-50\%$ and mean of $33\%$ given in Dolag \etal (2010), which matches
the $36\%$ measured for or cluster B35.

In addition to a simple surface brightness threshold, there is a
second commonly used definition of ICL measured from deep surface
photometric observations.  In this method, the cluster's galactic
light is modeled with idealized galaxy profiles and subtracted, with
the residual light defined as the ICL (\eg Gonzalez \etal 2005; Seigar
\etal 2007).  In particular, this method is commonly used on cD
galaxies, which traditionally have been defined as central galaxies
which display a ``double de Vaucouleurs'' profile, or an excess of
luminosity at large radius over the traditional de Vaucouleurs
profile, which is then fit by a second de Vaucouleurs profile (\eg de
Vaucouleurs 1953; Matthews \etal 1964; Oemler 1973; Feldmeier \etal
2002; Gonzalez \etal 2005).  The integrated light of this second,
outer profile is taken to be the ICL.  This method potentially
resolves the discrepancy between the theoretically favored binding
energy definition of ICL and our observationally tractable surface
brightness threshold method, whereby a surface brightness threshold
does not identify stripped material at the cluster's center as ICL,
since the fitted outer profile does indeed extend to the center of the
galaxy.  As expected, such analyses tend to yield higher ICL fractions
than an isophotal cut and similar to those from simulations employing
a binding energy ICL definition, with the ICL fraction typically at or
above $20-30\%$ (Seigar \etal 2007; Gonzalez \etal 2007; Zibetti \etal
2008).
 
While it is commonly assumed that the outer envelope is the product of
tidal stripping within the cluster environment in accordance with its
classification as ICL, it is unclear how well this profile truly
matches with the cluster's stripped material, especially in the
innermost regions which coincide with the inner, bright regions of the
cD itself.  Moreover, recent evidence suggests that the structure of
giant elliptical galaxies in massive clusters may not be as
well-defined as previously assumed (\eg Kormendy \etal 2009).  It has
been suggested that cD galaxies, with their extended ICL component,
may be best described as double de Vaucouleurs (\eg Gonzalez \etal
2005), de Vaucouleurs plus an outer exponential (\eg Seigar \etal
2007), or even a single S\'{e}rsic profile (\eg Kormendy \etal 2009,
Janowiecki \etal 2010).  Given the uncertainty of the true structure
of cD galaxies in clusters, it is extraordinarily difficult to
accurately subtract the galactic light using these parametric methods
in order to measure the ICL component in a robust and precise manner.
Additionally, fitting such galactic profiles is sensitive to a large
dynamic range in radii, including the galaxies' innermost regions.
Given our known issues with properly resolving the structure of
galactic cores, we have not been able to study such galaxy fitting
measurements in our simulated clusters along side our other ICL
measurement techniques.  Further progress in this area will likely
involve a more detailed understanding of the structure and formation
of elliptical galaxies, especially within the dense cluster
environment.

While the many varying definitions of ICL that have been employed to
date have been an impediment to a full census of the ICL in galaxy
clusters, the situation may be much improved in the future as our
understanding of each ICL detection method progresses.  In particular,
simple definitions with few free parameters, such as an isophotal
limit for observational data sets, are straightforward to implement
and may prove to be more robust and repeatable measurements, which
better allow direct comparisons between multiple studies than more
complex galaxy profile fitting methods.  For simulations of the ICL
component, we detailed in Section \ref{sec:iclbindingenergy} many of
the issues inherent in a binding energy definition of ICL with which
any algorithm must contend, either explicitly or implicitly.  A
density-based definition of ICL may lack the physically-motivated
elegance of a binding energy definition, but has the advantages that
it is simple to calculate and has a well-defined dependence on its
free parameters, making comparisons between multiple studies easier
and more robust.  Moreover, in order to be directly comparable to
observational data sets, simulation studies must choose metrics
relying on only observationally available data, such as the the
surface brightness threshold method detailed in Section
\ref{sec:iclsb}.

Although the ICL component contains a significant fraction of a
cluster's luminosity, simply parameterizing this content with a single
number, the ICL fraction, does not fully capture its utility or
intricacy.  The generation of ICL is dependent on the specific
dynamics of the cluster in a very complex manner, and the difficulty
in uniquely defining the ICL component only compounds this complexity.
For detailed observations of individual clusters the ICL's spatial
structure, kinematics, and stellar populations, along with its
quantity, can provide a more complete picture of the ICL component and
may be used to better relate the ICL to the cluster's dynamics.  In
particular, a number of observational studies have linked discrete
tidal streams of ICL to specific galactic interactions within the
cluster (\eg Trentham \& Mobasher 1998; Gregg \& West 1998; Adami
\etal 2005; Mihos \etal 2005; Rudick \etal 2010; Janowiecki \etal
2010).  These streams not only allow us to probe the specific
histories of the galaxies involved in the interactions, but R09 found
that while most ICL is initially formed in such streams, there is an
inverse correlation between the quantity of ICL in streams and the
cluster's dynamical age.  The stellar populations which comprise the
ICL can also be used to determine its origins.  For instance, most
studies of the broadband colors of the ICL component have determined
that its color is consistent with an old, metal-poor population
similar to the outskirts of the giant elliptical galaxies found in the
clusters (\eg Zibetti \etal 2005; Pierini \etal 2008; da Rocha \etal
2008; Rudick \etal 2010).  Furthermore, planetary nebulae found within
the ICL population provide a means for studying the ICL's kinematics
(\eg Arnaboldi \etal 2004; Gerhard \etal 2007; Doherty \etal 2009).
As sample sizes of these objects increase, they will provide an
unprecedented opportunity to study the dynamics of the cluster
environment, by providing a large number of probes with which to trace
the cluster's gravitational potential (\eg Willman \etal 2004;
Sommer-Larsen \etal 2005) and to trace the orbits and dynamics of any
ICL streams or substructures.  A cluster's ICL fraction is thus only
one of many tools available for investigating the ICL content of the
cluster and using it to trace the cluster's dynamical history.

\section{Summary}\label{sec:summary}
In this paper we have applied a number of ICL detection techniques to
a suite of simulated clusters designed expressly for the purposes of
studying the intracluster component.  Our ICL definitions include
techniques which are commonly used on both observational and
simulation data.  We have examined precisely how each defines the ICL
component, and studied the systematic differences between each
method.  Here we summarize our main conclusions:

\begin{itemize}

 \item The various ICL measurement techniques produce systematically
  different ICL fractions.  Measurements which rely exclusively on the
  position of the cluster luminosity, such as surface brightness or
  instantaneous density, tend to yield ICL fractions significantly
  less than methods which utilize time or velocity information, such
  as density history or binding energy.  For example, the ICL fraction
  of cluster C2 at \zz\ is $13.4\%$ as measured by surface brightness
  threshold and $24.5\%$ by the binging energy definition.
  Additionally, each ICL definition relies on one or more adjustable
  free parameters.  These parameters can change the measured ICL
  fractions by up to a factor of two, even within a single definition.

\item The ICL measurements made using each of our techniques agree
  quite well with those found in the literature using similar methods.
  Thus, the varying ranges for the ICL fraction found in different
  studies may be a manifestation of how each identifies the ICL in
  clusters which are in fact similar.

\item Our measurements of the ICL fraction in our clusters at \zz\
  range from $9-36\%$ using any method.  While this is not intended to
  put firm bounds on the range of ICL fractions found in \zz\
  clusters, it does demonstrate that all massive galaxy clusters are
  expected to have a non-negligible ICL component.

\item The majority of the discrepancies between the various ICL
  definitions are related to the separation of the central galaxy from
  the surrounding ICL.  The ICL is centrally concentrated in the
  cluster, often around a single central galaxy, and there is often no
  unambiguous transition from the galaxy's extended outer profile to
  the ICL.

\item The quantity of ICL tends to increase with time under all of
  our ICL definitions.  However, the ICL fraction does not grow at a
  constant rate and is related to the specific dynamical evolution of
  the cluster.  Although more dynamically advanced clusters will, on
  average, have higher ICL fractions, individual clusters at a single
  timepoint may deviate from this trend.

\item The fact that a cluster's ICL fraction can be defined in so many
  different ways, leading to such widely different results,
  underscores the fact that the quantity of ICL alone does not fully
  describe this important luminous component.  The morphology, stellar
  populations, and kinematics of the ICL each provide additional
  insight into its nature.  Only by combining the information from
  each can the ICL's potential to reveal the cluster's dynamical
  history be fully exploited.

\end{itemize}

\acknowledgments The authors would like to thank Heather L. Morrison,
Paul Harding, and Daniel S. Akerib for many useful suggestions and for
advice on defining the focus of this paper.  We also thank the
anonymous referee for several suggestions which improved this work.
This work utilized the computing resources of the CWRU ITS High
Performance Cluster, and we are indebted to its staff for their
patience and expertise.  C.S.R. appreciates support from the Jason
J. Nassau Graduate Fellowship Fund.  J.C.M. acknowledges research
support from the NSF through grants ASTR 0607526 and ASTR 0707793.
This research was supported in part by the National Science Foundation
through TeraGrid resources provided by PSC under grant numbers
TG-AST070001T and TG-AST070024.

\appendix
\section{Binding Energy Algorithm}
As described in Section \ref{sec:iclbindingenergy}, our algorithm to
calculate the binding energy of luminous particles to the cluster
galaxies works by measuring a spherical mass density profile for each
galaxy, and analytically calculating the galaxy's potential as a
function of radius using Equation \ref{eqn:ebind}.  In addition to the
mass density profile itself, Equation \ref{eqn:ebind} also requires
that we find the truncation radius of the galaxy, \rtrunc, or the
radius beyond which there is no more mass contributing to the galaxy's
gravitational potential.  Our procedures for calculating these
quantities are described in detail below.

\subsection{Mass Density Profiles}
The first step in determining the galaxies' mass profiles is to
identify individual galaxies, and measure their precise central
positions and velocities.  Fortunately, identifying galaxy cores is
relatively straightforward, as they are defined by dense
agglomerations of stellar particles.  We simply use a
friends-of-friends clustering algorithm with a linking length of 500 pc
to find dense groups of stellar particles and use their mean positions
and velocities as those of the galaxies.  The positions and velocities
of these galactic cores are insensitive to the choice of linking
length, over a wide range of reasonably small values.

Around each galaxy core, we need to create a spherical mass density
profile as a function of radius.  Doing so, however, is complicated by
two main factors.  First, galaxy halos may not be intrinsically
spherical, and we must therefore ``average'' over this non-sphericity.
Additionally, the mass profiles of neighboring galaxies may overlap
one another, whereas we wish to measure the mass profiles of galaxies
individually.  We address these two issues simultaneously by measuring
a galaxy's mass density profile independently in eight octants around
the galaxy, in logarithmically spaced radial bins.  We see the
non-sphericity of the halo in the scatter of the density values
between different octants.  We find that this density scatter is
approximated by a log-normal distribution, and thus take the density
in a given radial bin to be the mean of the logarithm of the density
values in each octant.  Meanwhile, neighboring galaxies are easily
identified as sharp peaks in the density profiles, which usually are
present in only one or two octants.  We can therefore minimize the
impact of such galaxies by excluding the two highest density values at
every radius.  For isolated halos which do not have neighbors, we find
that restricting our density profile to the six least dense octants
does not significantly alter the galaxy's density profile or
gravitational potential.

\begin{figure*}
\plotone{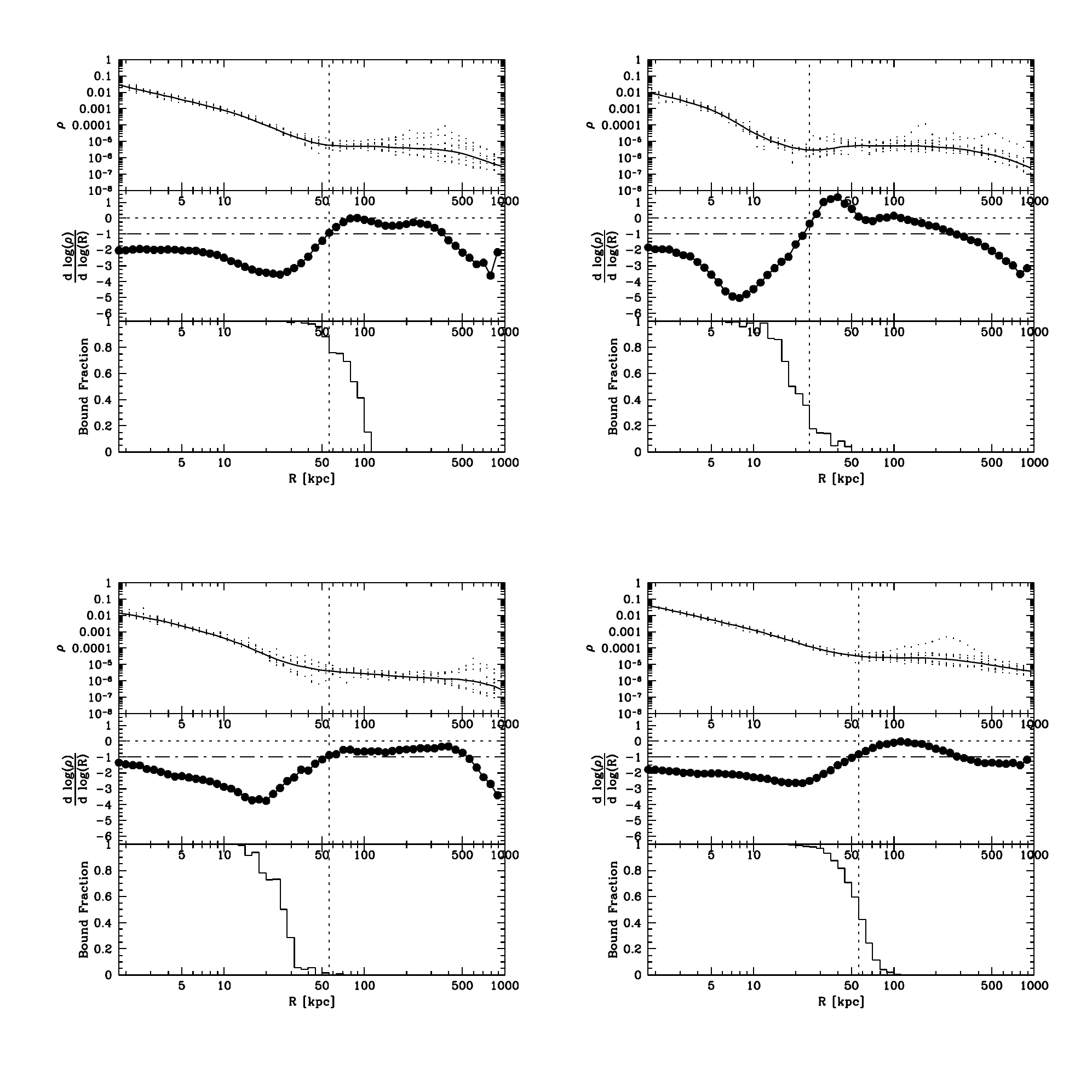}
\caption{ The density profiles of four typical satellite galaxies, each
  from a different cluster at \zz.  \emph{Top}: Dots show the raw
  densities in each of the eight octants around the galaxy, while the
  solid line shows the final smoothed density profile.  \emph{Middle}:
  The \drdr\ final smoothed \drdr\ profile of the galaxy.  The dotted
  line marks \drdr$=0$ while the dot-dash line marks \drdr$=-1$.
  \emph{Bottom}: The fraction of luminous particles at each radius
  which are bound to the galaxy.  The vertical dotted line running
  through all three plots shows the truncation radius of the galaxy.
  \label{fig:example_profiles}}
\end{figure*}

The final step in creating the mass density profile of each galaxy is
to smooth the profile.  Smoothing is necessary in order to reduce two
primary effects: 1) shot noise caused by the discreteness of the
particles used to measure the density (because of our multi-resolution
simulation approach, the outskirts of many galaxies are sampled by
only a small number of relatively high-mass particles), and 2)
short-lived variations in the mass distribution of the galaxies which
may affect the density distribution at a particular time, but do not
have an impact on the long-term binding state of stellar particles.
Therefore, in order to effectively identify features in the density
profiles, they are smoothed using a simple boxcar smoothing kernel.
Additionally, because the analyses below rely on identifying features
in a galaxy's \drdr\ profile, we then smooth this derivative profile,
and recreate the final density profile from the central density and
the smoothed derivative to ensure consistency.  Again, testing has
shown that the minor adjustments to the density profile created by
this smoothing process have very little effect on the final binding
energy profiles of the galaxies.  Several example galaxy profiles are
shown in Figure \ref{fig:example_profiles}.  The top plots of the
figure show the galaxies' density profiles, with the small points
showing the raw density measurements in the eight octants, while the
solid line shows the final smoothed profile.  The smoothed \drdr\
profiles are shown in the middle plots.

\subsection{Truncation Radius}
A galaxy's truncation radius, \rtrunc, is essentially the ``edge''
of each galaxy, or the radius beyond which there is no mass which
contributes to the galaxy's gravitational potential.  For the majority
of galaxies within a cluster halo which do not reside at the center of
the cluster, known as satellite galaxies, we essentially want to find
the radius at which the background density of the cluster itself
begins to dominate the galaxies' mass distribution.

\begin{figure}
\plotone{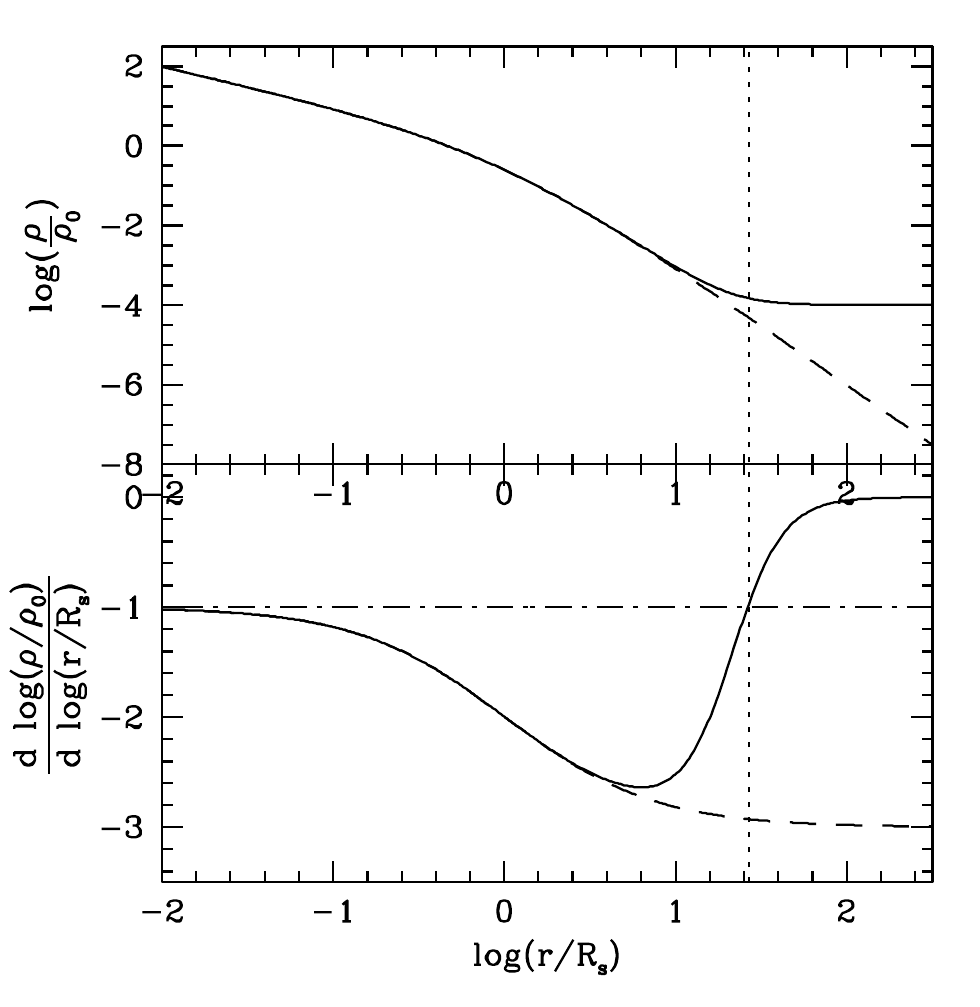}
\caption{An idealized NFW profile, with unit scale radius, $R_{s}$, and
  characteristic density, $\rho_{0}$.  \emph{Dashed Lines}: Pure NFW
  profile, indicative of an isolated galaxy.  \emph{Solid Lines}: NFW
  profile embedded in a constant density background, indicative of a
  galaxy within the cluster environment.  \emph{Top}: The galaxies'
  density profiles.  \emph{Bottom}: The galaxies' \drdr\ profiles.
  The dotted vertical line indicates the truncation radius while the
  dot-dash line marks \drdr$=-1$.
\label{fig:nfw}}
\end{figure}

We illustrate the main features of our approach with the highly
idealized example seen in Figure \ref{fig:nfw}.  Figure \ref{fig:nfw}
shows two galaxy mass density profiles, a pure NFW profile --- an
isolated galaxy --- and the same NFW halo embedded within a constant
density background --- a galaxy residing within a massive cluster
potential.  In the inner regions, the two halos are essentially
identical.  However, whereas the pure NFW halo continues to drop in
density even at the largest radii, the embedded halo levels off at the
background density.  The different behavior of these two profiles can
be very clearly seen in the plots of their derivatives, \drdr, also
shown in Figure \ref{fig:nfw}.  Whereas the slope of the pure NFW halo
decreases from $-1$ in the inner regions to $-3$ at the largest radii,
the embedded halo's slope reaches a minimum and then increases to an
asymptotic value of $0$.  This derivative curve clearly indicates that
as the radius increases, the constant background density begins to
dominate over the profile of the galaxy itself, and we can use the
features of this curve to define the truncation radius.  We have
chosen to define the truncation radius as the point where the \drdr\
slope exceeds $-1$, which is marked with a dotted line in Figure
\ref{fig:nfw}.  From Figure \ref{fig:nfw}, it is clear that
qualitatively this is the radius at which the density profile is
leveling out to the background density.  Importantly, while this
analysis clearly demonstrates the utility of our method for NFW galaxy
halos, it is not dependent on the specific properties of the NFW
profile, and is equally effective on any similar declining galaxy mass
density profile.

The middle panels of Figure \ref{fig:example_profiles} show the
\drdr\ curves for our example simulation galaxies.  The galaxies in
Figure \ref{fig:example_profiles} show the same qualitative behavior
as our idealized example in Figure \ref{fig:nfw}, where the slope of
the density profile decreases in the inner regions, and then reaches a
minimum before increasing to \drdr$\approx 0$ at large radii.
Extensive testing has shown that defining the truncation radius to be
the point where the derivative is $\ge-1$ is a robust measurement for
the vast majority of simulation galaxies.  The examples in Figure
\ref{fig:example_profiles} demonstrate that this metric delineates the
qualitative transition where the mass density profiles of the galaxies
begin to level off and be dominated by the background cluster density.
For galaxies of the same stellar mass, those which are at larger
cluster-centric radii will generally have larger truncation radii than
galaxies near the cluster center, due to the fact that the cluster's
background density decreases with radius.  We illustrate this point
with two disk galaxies of similar mass from cluster C2, but at two
different cluster-centric radii.  While each galaxy has a stellar mass
of $\approx3\times10^{10}$ \Msun and a half-light radius of $\approx3$
kpc, the galaxy which lies 850 kpc from the cluster center has a
truncation radius of 140 kpc, while the galaxy at cluster-centric
radius of 264 kpc has a truncation radius of only 40 kpc.  The bottom
plots in Figure \ref{fig:example_profiles} show the fraction of the
stellar particles bound to each galaxy as a function of radius.  Each
galaxy shows a similar pattern, where almost all the stellar mass at
small radii is bound, and there is a fairly sharp transition at or
near truncation radius where the vast majority of stellar mass becomes
unbound.

\begin{figure*}
\plotone{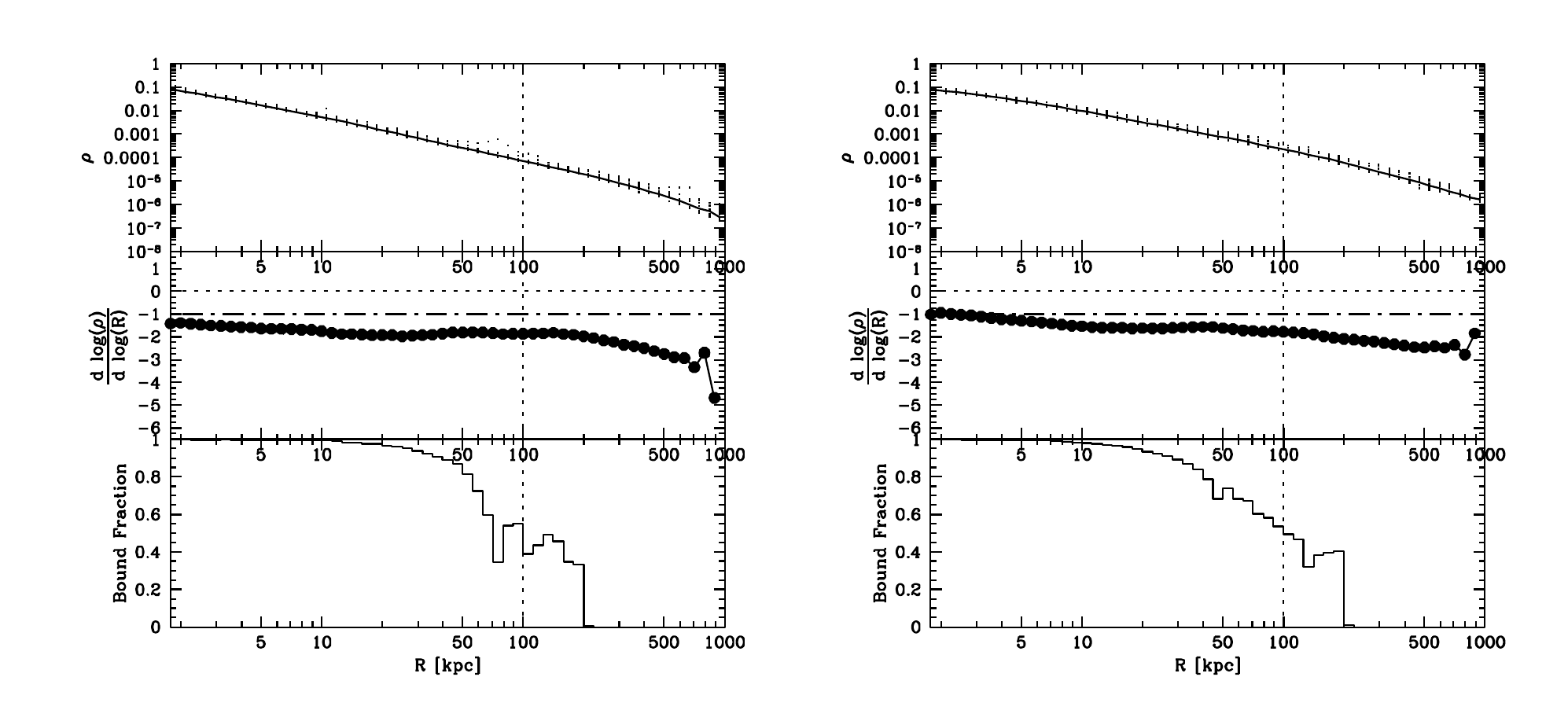}
\caption{The density profiles of the massive central galaxies from
  clusters C2 (left) and B35 (right) at \zz.  Plots are
  the same as in Figure \ref{fig:example_profiles}.
\label{fig:cd_profiles}}
\end{figure*}

There is one particularly important instance, however, when the
assumptions used in the algorithm described above break down.  For
galaxies which sit very near the center of the cluster, especially cD
galaxies, the paradigm of a galactic halo embedded within a locally
constant density background does not apply.  Instead, the mass density
profiles of these galaxies are essentially indistinguishable from
those of the cluster as a whole.  Figure \ref{fig:cd_profiles} shows
the density profiles of two central galaxies, from clusters C2 and B35
at \zz.  As expected, these density profiles look like cluster-mass
NFW profiles and show no indication of leveling off at radii up to 1
Mpc.  Fundamentally, these galaxies do not have dark halos which are
independent of the overall cluster potentials, and thus any mass bound
to the cluster is then bound to the the central galaxies.  This, of
course, would completely obviate our definition of ICL, in which ICL
is luminous particles bound to the cluster but no galaxy within the
cluster, since \emph{all} luminous particles would be bound to the
central galaxy.

Thus, in defining \rtrunc, central galaxies must be treated as a
special case.  This issue is compounded by the fact that while central
galaxies in massive clusters are readily identified by eye, there are
other scenarios in which the same conditions hold.  For instance,
galaxies at the center of group-mass halos will also be
indistinguishable from the overall group mass distribution.  This
makes it extremely difficult to implement such a binding energy
measurement technique at higher redshifts, when the cluster mass is
contained in a number of smaller groups which later merge to form the
$z=0$ cluster.  In order to efficiently automate the calculations, we
have simply imposed a maximum truncation radius (described below), and
thus it is not necessary to manually specify any central galaxies.

\subsubsection{Binding Energy Free Parameters}
Given the algorithm described above, the major free parameter which
can be varied is the maximum truncation radius.  This parameter
primarily affects the binding state of particles due to the change in
potential of the clusters' central galaxies.  Our preferred value for
the maximum truncation radius is 100 kpc.  While several of the
largest satellite galaxies have truncation radii based on their
density profiles which are slightly larger than this limit, up to
$\approx 200$ kpc, this has very little effect on the total measured
ICL in the cluster.  We illustrate this effect with a series of tests
on cluster C2, where we calculate the luminosity bound to the
satellite galaxies (\ie all galaxies except for the central galaxy)
When the maximum truncation radius is set to 100 kpc, 42.3\% of the
luminosity is bound to the satellite galaxies, while when the maximum
radius is 200 kpc 44.9\% of the cluster luminosity is bound to these
galaxies.  Moreover, changing the algorithm so that \rtrunc\ is
defined as the point where the \drdr\ curve reaches a minimum
(\ddrddr$=0$), has a similarly small effect on the satellite galaxies.
This analysis demonstrates that the gravitational potentials of
satellite galaxies are fairly readily defined, and there is a
relatively small amount of material that is only marginally
bound/unbound from these galaxies.

The cluster ICL fraction, is, however, very sensitive to the
truncation radius of the \emph{central} galaxy.  Whereas the maximum
truncation radius only marginally affects a small number of satellite
galaxies, this maximum radius will always be the truncation radius of
the central galaxy, and has a very large effect on its gravitational
binding energy.  We again demonstrate with tests on cluster C2, this
time running the binding energy algorithm on all galaxies, including
the central galaxy.  With a truncation radius of 100 kpc, 20.0\% of
the cluster luminosity is unbound to any galaxy, whereas only 9.5\% of
the cluster luminosity is unbound when \rtrunc\ is set to 200 kpc.
From the previous tests, we know that the vast majority of this change
comes from the central galaxy.  From Figure \ref{fig:cd_profiles}, it
is clear that the mass density of the central galaxy in this radial
range is very high, and thus small changes in the truncation radius
will have a large effect on the galaxy's binding energy.  However, as
there are no clear features in the mass density or luminosity profiles
of the galaxy to use in defining the truncation radius, any value will
be somewhat subjective and arbitrary.  Moreover, because the cluster's
luminosity is so centrally concentrated (see Section
\ref{sec:iclfeatures}), the binding energy of the central galaxy
has a huge effect on the total ICL fraction of the cluster.

The bottom plots of Figure \ref{fig:cd_profiles} show the fraction of
the clusters' luminous mass bound to the central galaxies as a
function of radius, similar to Figure \ref{fig:example_profiles}.
These galaxies do not show nearly as sharp a transition from bound to
unbound near the truncation radius.  Furthermore, a substantial
fraction of the clusters' luminosity is bound to the central galaxies,
even at very large radii.  We have therefore limited the binding
energy calculations such that stellar particles beyond twice the
truncation radius cannot be bound.  This restriction has almost no
effect on the satellite galaxies, but for cluster C2 the ICL fraction
increases by $\approx 4\%$ due to particles no longer being bound to
the central galaxy.

\end{document}